\documentclass[conference]{IEEEtran}
\ifCLASSINFOpdf
\else
\fi
\usepackage{graphics}
\usepackage{graphicx}
\usepackage{enumerate}
\usepackage{epstopdf}
\usepackage{subfigure}
\usepackage{amsmath,amsfonts}
\usepackage{bm}
\usepackage{enumerate}
\usepackage{multirow}
\usepackage{url}
\usepackage{xcolor}

\def\hlinewd#1{%
  \noalign{\ifnum0=`}\fi\hrule \@height #1 \futurelet
   \reserved@a\@xhline}
\makeatother

\newcommand{\tabincell}[2]{\begin{tabular}{@{}#1@{}}#2\end{tabular}}%

\newcommand{\new}{\textcolor{black}}
\newcommand{\hank}{\textcolor{black}}
\newcommand{\ruby}{\textcolor{black}}

\begin{document}
%

\title{Implicit Smartphone User Authentication with Sensors and Contextual Machine Learning}





%
\author{\IEEEauthorblockN{Wei-Han Lee and
Ruby B. Lee}
\IEEEauthorblockA{Princeton University\\ 
Email: \{weihanl, rblee@princeton.edu \}}
}


%

\maketitle

\begin{abstract}
Authentication of smartphone users is important because a lot of sensitive data is stored in the smartphone and the smartphone is also used to access various cloud data and services. However, smartphones are easily stolen or co-opted by an attacker. Beyond the initial login, it is highly desirable to re-authenticate end-users who are continuing to access security-critical services and data. Hence, this paper proposes a novel authentication system for implicit, continuous authentication of the smartphone user based on behavioral characteristics, by leveraging the sensors already ubiquitously built into smartphones. We propose novel context-based authentication models to differentiate the legitimate smartphone owner versus other users. We systematically show how to achieve high authentication accuracy with different design alternatives in sensor and feature selection, machine learning techniques, context detection and multiple devices. Our system can achieve excellent authentication performance with $\bm{98.1\%}$ accuracy with negligible system overhead and less than $\bm{2.4\%}$ battery consumption.
\end{abstract}
\section{Introduction}
Increasing amounts of private and sensitive information are stored in our smartphones. $92.8\%$ of Android smartphone users store private information in their smartphones \cite{kim2015analyzing, achara2013mobilitics}. Smartphones have also become personal computing platforms for users to access cloud services, e.g., e-banking and online social networks. Hence, smartphones are very attractive targets for attackers to get access to personal and valuable information. User authentication is essential to prevent the privacy, confidentiality and integrity breaches possible through attacks on the smartphone. 

Current login mechanisms use explicit authentication, which requires the user's participation, e.g., passwords and fingerprints. \hank{Iris scanning \cite{qi2008iris} and facial recognition \cite{xi2012mobile,niinuma2010soft}} can also be used for explicit authentication. However, re-authentication to access very sensitive information via explicit authentication mechanisms is not convenient~\cite{consumer} for smartphone users. Hence, after the user passes the initial authentication, the system does not authenticate the user again. This creates a significant risk for adversaries to take control of the users' smartphones, after the legitimate users' initial login. This enables the adversaries to access proprietary or sensitive data and services, whether stored in the cloud or in the mobile device itself.

To protect smartphone data and cloud-based services from adversaries who masquerade as legitimate users, we propose a secure re-authentication system, which is both \textit{implicit} and \textit{continuous}. An implicit authentication method does not rely on the direct involvement of the user, but is closely related to her behavior recorded by the smartphone's built-in hardware, e.g., sensors, GPS and touchscreen. An implicitly continuous re-authentication method should keep authenticating the user, in addition to the initial login authentication, without interrupting users. This can detect an adversary once he gets control of the smartphone and can prevent him from accessing sensitive data or services via smartphones, or inside smartphones.

Our system, called SmarterYou, exploits one of the most important differences between personal computers and smartphones: a variety of sensors built into the smartphone, such as the accelerometer and gyroscope. SmarterYou also exploits the increasing number of wearable devices with Bluetooth connectivity and multiple sensors, e.g., smartwatches.

SmarterYou has the following advantages compared with previous smartphone authentication methods: (1) \new{Instead of the explicit one-time authentication on log-in, e.g., using passwords, fingerprints or touchscreen patterns~\hank{\cite{de2012touch, clarke2007authenticating}}, SmarterYou enables implicit, continuous authentication as a background service, when the users use smartphones. This can also be used in addition to the explicit authentication methods.} (2) We do not require user's permissions. Many past approaches require the user's permission to get access to the hardware in the smartphone, e.g., GPS~\cite{cc4} and microphone~\cite{riva2012progressive}. Access to these hardware require permission because they contain private information of the user. (e.g., her location and phone conversations). (3) Some past work had high authentication errors~\cite{cc2, mantyjarvi2005icassp}. Our approach can have accuracy up to $98.1\%$. (4) Many approaches utilize the touchscreen to analyze user's writing or sliding patterns. However, the touchscreen information may leak out sensitive information, e.g., passwords or PINs \hank{\cite{cc8, aviv2010smudge}}. (5) Many past approaches only work under some specific context~\cite{conti2011swing, cc7, cc5, mantyjarvi2005icassp, okumura2006study}. In SmarterYou, we utilize multiple contexts to improve authentication accuracy, and also design a context detection method that is user-agnostic. 

\begin{table*}\label{table:related}\scriptsize
\caption{Comparison of our method with other implicit authentication (if the information is given in the paper cited, otherwise it is shown as n.a. (not available)) for authentication accuracy, false accept rate (FAR) and false reject rate (FRR).}\centering
\begin{tabular}{|l|c|c|c|c|c|} \hline
            			& Modality   & \multicolumn{3}{|c|}{Performance} & \# of Users \\ \cline{3-5}
            			&             & Accuracy & FAR & FRR  &\\ \hline
\cite{cc5} Trojahn et al. 2013				  & Touchscreen & n.a.   & $11\%$ & $16\%$ & 18 \\ \hline
\cite{frank2013touchalytics} Frank et al. 2013& Touchscreen & $96\%$   & n.a. & n.a. & 41 \\ \hline
\cite{cc6} Li et al.                      2013& Touchscreen & $95.7\%$ & n.a. & n.a. & 75\\ \hline
\cite{feng2012continuous} Feng et al.     2012& Touchscreen \& accelerometer \& gyroscope & n.a. & $4.66\%$ & $0.13\%$ & 40\\ \hline
\cite{xu2014towards} Xu et al.           2014 & Touchscreen & $>90\%$ & n.a. & n.a. & 31\\ \hline
\cite{zheng2014you} Zheng et al.         2014 & Touchscreen \& accelerometer & $96.35\%$ & n.a & n.a. & 80 \\ \hline
\cite{conti2011swing} Conti et al. 		 2011 & accelerometer \& orientation & n.a. & $4.44\%$ & $9.33\%$ & 10 \\ \hline
\cite{cc3} Kayacik et al.                2014 & accelerometer \& orientation \& magnetometer \& light & n.a. & n.a. & n.a. & 4 \\ \hline
\cite{cc2} Zhu et al.					 2013 & accelerometer \& orientation \& magnetometer & $75\%$ & n.a. & n.a. & 20 \\ \hline
\cite{cc7} Nickel et al. 				 2012 & accelerometer & n.a. & $3.97\%$ & $22.22\%$ & 20 \\ \hline
\cite{lee2015icissp} Lee et al.			 2015 & accelerometer \& orientation \& magnetometer & $90\%$ & n.a. & n.a. & 4 \\ \hline
\cite{yang2015unlocking} Yang et al.	 2015 & accelerometer & n.a. & $15\%$ & $10\%$ & 200 \\ \hline
\cite{cc4}  Buthpitiya et al.            2011 & GPS           & $86.6\%$ & n.a. & n.a. & 30\\ \hline
SmarterYou (this paper)  2017                 & accelerometer \& gyroscope & $98.1\%$ & $2.8\%$ & $0.9\%$ & 35 \\ \hline
\end{tabular}
\end{table*}

\hank{In this paper, we utilize context detection techniques and multiple mobile devices to achieve accurate authentication performance stealthily, efficiently, and continuously. 
Also, we protect cloud-customers' services and data from malicious end-users using smartphone sensors. We also provide a systematic evaluation of the design alternatives for our system, in terms of sensors, features, contexts, multiple devices and machine learning algorithms. Our key contributions are:}

\begin{enumerate}[$\bullet$]
\item Design of an implicit authentication system, SmarterYou, by combining a user's information recorded in the smartphone and wearable devices. Our system continuously monitors a user's behavior and re-authenticates the user in an accurate, efficient, and stealthy manner.
\item Design of a user-agnostic context detection approach to differentiate various usage contexts of the user. We determine the minimum number of contexts that give the best improvement in authentication performance.
\item Design and evaluation of alternatives for all aspects of an efficient authentication method based on sensor measurements used as behavioral patterns. We consider the minimum number of sensors for high authentication accuracy, the best features in both time and frequency domains, the benefit of using multiple devices with sensors, the advantages of user context-specific authentication models, and alternative machine learning algorithms. To the best of our knowledge, this is the first systematic evaluation of design alternatives for sensor-based user authentication.
\item SmarterYou also provides automatic and continuous retraining if the user's behavioral pattern changes over time. We also evaluate the performance overhead and battery consumption of the system.
Our approach can achieve high authentication accuracy up to $98.1\%$ with negligible system overhead and less than $2.4\%$ battery consumption.
\end{enumerate}
\begin{figure*}[t]
\centerline{\includegraphics[width=0.7\linewidth]{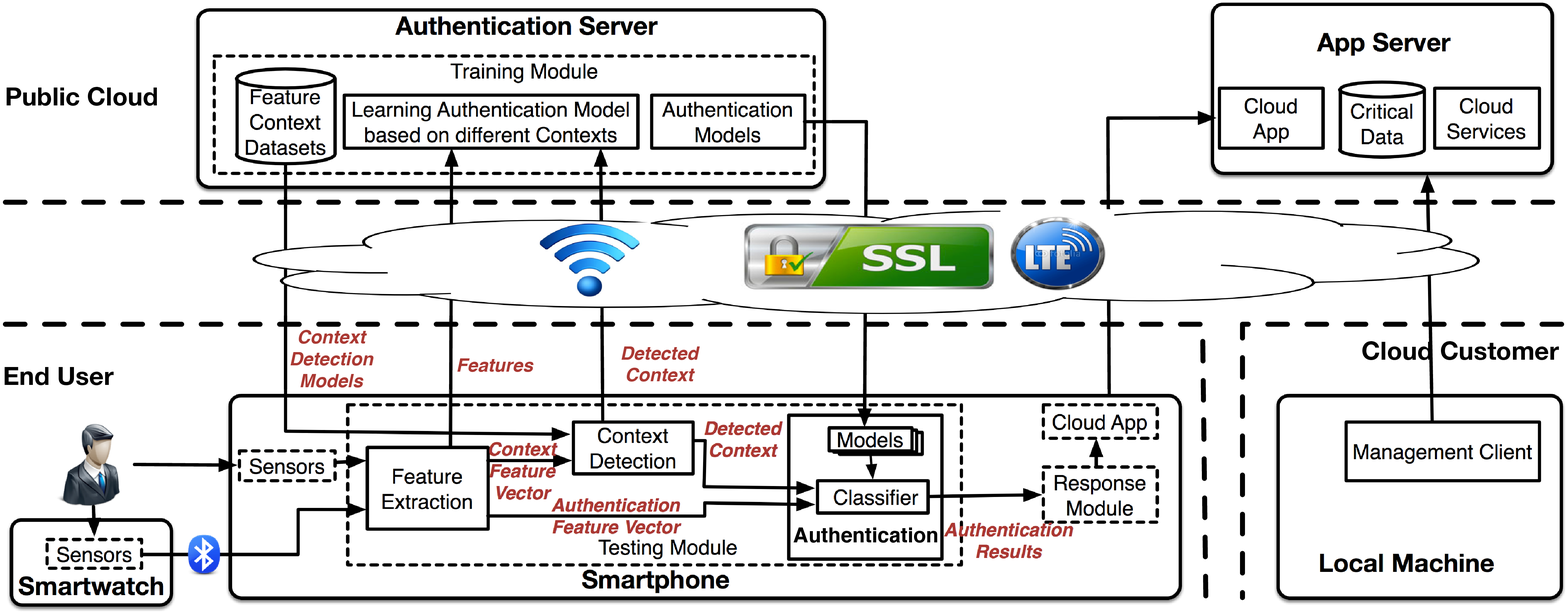}}
\caption{SmarterYou architecture including the cloud-based training module and smartphone-based authentication module}
\label{fig:shadow}
\end{figure*}
\section{Background and Related Work}\label{sec:related}
Traditional authentication approaches are based on possession of secret information, such as passwords. Also, physiological biometrics based approaches make use of distinct personal features, such as fingerprint or iris patterns. Recently, behavior-based authentication utilize the distinct behavior of users. 

There are many different physiological biometrics for authentication, such as face patterns \hank{\cite{niinuma2010soft}}, fingerprints \cite{hong1998pami}, and iris patterns \cite{qi2008iris}. However, physiology-based authentication requires user participation in the authentication. Thus, they are more useful for initial login authentication instead of implicit and continuous authentication and re-authentication.

Behavior based authentication assumes that people have distinct, mostly stable, patterns for a certain behavior, such as gesture pattern \cite{cc5}, gait \cite{cc7} and GPS patterns \cite{cc4}. Behavior-based authentication exploits users' behavioral patterns to authenticate a user's identity. Below we review past work in this area and summarize them in Table \ref{table:related}.

\noindent{\bf Touchscreen-based Smartphone Authentication.}

Trojahn et al. \cite{cc5} developed a mixture of a keystroke-based and handwriting-based mechanisms to realize authentication through the touchscreen sensor. Their approach has achieved $11\%$ false accept rate (FAR) and $16\%$ false reject rate (FRR).
Frank et al. \cite{frank2013touchalytics} utilize 22 analytic features from sliding traces to differentiate users. Their result can achieve $4\%$ equal error rate. 
Li et al. \cite{cc6} exploited five basic movements (sliding up, down, right, left and tapping) on the touchscreen and their related combinations as the user's behavioral pattern features, to perform authentication. Their result shows that sliding up can achieve the best accuracy of $95.7\%$. 
Feng et al. \cite{feng2012continuous} utilize touchscreen with sensor gloves to record the fine-grained information of gestures. They can achieve up to $4.66\%$ FAR and $0.13\%$ FRR. 
Xu et al. \cite{xu2014towards} combine the slide, keystroke, handwriting and pinch to authenticate the user. 
Zheng et al. \cite{zheng2014you} combine the accelerometer with the touchscreen to authenticate a user when the user is entering her PIN. 

Touchscreen-based authentication can achieve high accuracy. However, Serwadda et al. \cite{serwadda2013kids} showed that gesture styles could be observed and replicated automatically. Also, the touchscreen information contains sensitive information, e.g., the attacker may use the touchscreen information to find out the user's passwords \cite{cc8}.

\noindent{\bf Sensor-based Smartphone Authentication.}

Conti et al. \cite{conti2011swing} proposed to authenticate a user using the arm movement patterns, sensed by the accelerometer and orientation sensor, while the user is making a phone call. Their method achieved $4.4\%$ FAR and $9.3\%$ FRR.
Kayacik et al. \cite{cc3} proposed a light-weight, and temporally \& spatially aware user behavioral model for user authentication based on both hard and soft sensors. However, they did not quantitatively show their authentication performance.
SenSec \cite{cc2} constantly collects data from the accelerometer, gyroscope and magnetometer, to construct gesture models while the user is using the device. SenSec has shown it can achieve 75\% accuracy in identifying owners. 
Nickel et al. \cite{cc7} proposed an accelerometer-based behavior recognition method to authenticate a smartphone user through the $k$-NN algorithm. They can achieve $3.97\%$ FAR and $22.22\%$ FRR.
Lee et al. \cite{lee2015icissp} showed that using more sensors can improve the authentication performance. They monitored the users' living patterns and utilized SVM as a classifier for user authentication. Their result achieves $90\%$ accuracy.
Yang et al. \cite{yang2015unlocking} propose a hand waving biometric-based authentication method that utilise users' waving patterns for locking and unlocking the smartphone by using the accelerometer. They can achieve $15\%$ FAR and $10\%$ FRR on average.
In \cite{cc4}, a geo-based authentication is proposed for modeling a user's mobility pattern. They use the GPS sensor to demonstrate that the system could detect abnormal activities (e.g., a phone being stolen) by analyzing a user's location history, and they can achieve $86.6\%$ accuracy. However, the GPS information is sensitive, thus its use requires explicit user permission. 

Different from these past methods, our SmarterYou system has broader contexts and the highest authentication accuracy ($98.1\%$), with low computational complexity. 

\noindent{\bf Continuous and Context-based Authentication.}

Riva et al. \cite{riva2012progressive} built a prototype to use face recognition, proximity, phone placement, and voice recognition to progressively authenticate a user. However, their objective is to decide when to authenticate the user and is thus orthogonal to our setting. Their prototype has a $42\%$ reduction in requested explicit authentication, but this was conducted with 9 users only. Their scheme also requires access to sensors that need users' permissions, limiting their applicability for implicit, continuous authentication proposed in our system.

Existing continuous authentication approaches~\cite{cc5,cc7,conti2011swing} focused on a specific usage context and would fail if the attacker who steals the smartphone does not perform under the specific usage context. In contrast, our system can automatically detect a context in a user-agnostic manner and can continuously authenticate a user based on various authentication models. That is, our system can authenticate the users without requiring any specific usage context, making it more applicable in real world scenarios. 

\noindent{\bf Authentication with Wearable Devices.}

Recently, wearable devices have emerged in our daily lives. However, limited research has been done on authenticating users by these wearable devices. Mare et al.~\cite {mare2014zebra} proposed ZEBRA which is a bilateral recurring authentication method. The signals sent from a bracelet worn on the user's wrist are correlated with the terminal's operations to confirm the continued presence of the user if the two movements correlate according to a few coarse-grained actions. 
To the best of our knowledge, there is no smartphone authentication research proposed in the literature that combines a wearable smartwatch with a smartphone to authenticate a user, as we do.
\section{Threat model and Assumptions} \label{sec:threat}
We consider an attacker who has physical access to a smartphone. The smartphone may even have passed an initial explicit login authentication, giving the attacker opportunity to access secure or private information on the phone and in the cloud using the phone. Confidentiality, integrity, authentication and privacy breaches are considered.

Wearable devices are gaining popularity, e.g., smartwatches and fitbits. They also contain many sensors e.g., accelerometer, gyroscope, ambient light and heartbeat sensors, and can communicate with smartphones via Bluetooth. We assume each smartwatch (and smartphone) is associated with one owner/user and that users do not share their smartwatches (and smartphones). We assume the communication between the smartwatch and smartphone is secure. We do not assume that users always have their smartwatch with them, so authentication based on smartphone alone is in scope.

While network access is required for authentication model training, or retraining after behavioral drift, network access is not required for user authentication (testing) when the smartphone is being used.

\section{System Design} \label{sec:design} 
\subsection{Architecture Overview}\label{archi}
Figure~\ref{fig:shadow} shows the proposed SmarterYou architecture. It includes three hardware devices: the user-owned wearable device (e.g., smartwatch), the smartphone, and the authentication server in the cloud.

\subsubsection{Wearable IoT device} 
In SmarterYou, we consider a two-device authentication configuration, which includes a smartphone and a user-owned wearable device. We use a smartwatch as an example, but other types of wearable devices, e.g., health sensors, can also be applied to SmarterYou. SmarterYou is designed for implicit authentication on the smartphone, where the smartwatch serves as important auxiliary information for improving authentication accuracy. The smartwatch keeps monitoring a user's raw sensors' data and sends the information to the smartphone via Bluetooth. Our system works if only the smartphone is present, but we will show that it works even better if the smartwatch is also present.

\subsubsection{Smartphone}
Similar to the smartwatch, the smartphone also monitors the user's sensor data. It runs the authentication testing module as a background service in the smartphone. In the testing module, the feature extraction component receives the sensor data from the smartphone and smartwatch. Then it extracts fine-grained time-frequency features from the raw data, and forms two feature vectors: the context feature vector and the authentication feature vector, and feeds them into the context detection component and the authentication component, respectively. The context detection component decides which context the user is in and sends the detected context to the authentication component. 

The authentication component consists of a classifier and multiple authentication models. The classification algorithm we selected is the kernel ridge regression (KRR) algorithm \cite{suykens2002krr}, but other machine learning algorithms can also be used. An authentication model is a file containing parameters for the classification algorithm and determines the classifier's functionality. Using different authentication models for different contexts, the classifier can authenticate the user based on the authentication feature vector under different contexts. When a detected context and an authentication feature vector is fed in, the classifier chooses the corresponding authentication model and makes a classification. 

When the classifier in the authentication component generates the authentication results, it sends these results to the Response Module. If the authentication results indicate the user is legitimate, the Response Module will allow the user to use the cloud apps to access the critical data or cloud services in the app server. Otherwise, the Response Module can either lock the smartphone or refuse accesses to security-critical data, or perform further checking. Our system can be used with existing explicit authentication methods, e.g., passwords or fingerprints. If the attacker is locked out, the system requires explicit authentication.


\subsubsection{Authentication Server} 
SmarterYou includes a training module, which is deployed in the Authentication Server in the cloud, because it requires significant computation and must consider the privacy of the training data set, which includes data from other users. When a legitimate user first enrolls in the system, she downloads the context detection model from the Authentication Server and then the system keeps collecting the legitimate user's authentication feature vectors and detected contexts for training the authentication models. Our system deploys a trusted Authentication cloud server to collect sensors' data from all the participating legitimate users. To protect a legitimate user's privacy, the users' data are anonymized. In this way, a user's training module can use other users' feature data but has no way to know the other users' identities. The training module uses the legitimate user's authentication feature vectors and other people's authentication feature vectors in the training algorithm to obtain the authentication models based on different contexts. After training, the authentication models are downloaded to the smartphone. The training module does not participate in the authentication testing process and is only needed for retraining when the device recognizes a user's behavioral drift, which is done online and automatically. Therefore, our system does not pose a high requirement on the communication delay between the smartphone and the Authentication Server.

\subsection{System Operation}
SmarterYou is based on the observation that users' behavioral patterns are different from person to person, and vary under different usage contexts, when they use smartphones and smartwatches. 
Instead of authenticating the user with one unified model as in~\cite{conti2011swing,cc7,cc5,mantyjarvi2005icassp,okumura2006study,lee2017secure},
it is better to explore different finer-grained models to authenticate the user based on different usage contexts. For example, using a user's walking behavioral model to authenticate the same user who is sitting while using the smartphone is obviously not accurate. In Table~\ref{table:krr}, we show that considering contexts provides better accuracy. To be applicable in real world scenarios, we assume that the context information is user-agnostic: we can detect the context of the current user prior to authenticating her (as validated in Section \ref{sec:contextDetection}). Under each context, each user has distinct behavioral characteristics. SmarterYou utilizes such characteristics to implicitly authenticate the users. Our system can be used with other context detection methods \cite{chen2015designing, kern2003multi}. Context detection is an interesting research area e.g., Chen et al. \cite{chen2015designing} show that they can achieve up to $99\%$ accuracy in context detection. In this paper, we show that by considering even simple contexts, we can improve the authentication accuracy significantly. More contexts, appropriately chosen, may further improve the authentication accuracy.

There are two phases for learning and classifying the user's behavioral pattern: \textit{enrollment phase} and \textit{continuous authentication phase}. 

\noindent{\bf{Enrollment Phase}}: Initially, the system must be trained in an enrollment phase. When users want to use the apps in the smartphone to access sensitive data or cloud services, the system starts to monitor the sensors and extract particular features from the sensors' data and label them with a context based on the context detection approach in Section~\ref{sec:contextDetection}. This process continues and the data should be stored in a protected buffer in the smartphone until the distribution of the collected features converges to an equilibrium, \hank{which means the size of data can provide enough information to build a user's profile. This is about $800$ measurements for our method, as shown in Section~\ref{datasize}.} At this time, one can assume that 1) the user got used to her device and her device-specific `sensor-behavior' no longer changes, and 2) the system has observed sufficient information to have a stable estimate of the true underlying behavioral pattern of that user. The system can now train the authentication classifiers under various contexts and switch to the continuous authentication phase.


\noindent{\bf{Continuous Authentication Phase}}: Once the authentication classifiers are trained and sent to the smartphone, the smartphone can start the authentication phase. This is done only in the smartphone, so network availability is not required. Based on the sensor data, SmarterYou first decides which context the user is in and then uses the authentication classifier for the detected context. The authentication classifier then decides whether these sensors' data are coming from the legitimate user. The authentication classifier can also be automatically updated when the legitimate user's behavioral pattern changes with time.

\noindent{\bf{Post-Authentication}}: If the authentication feature vector is authenticated as coming from the legitimate user, this testing passes and the user can keep accessing the sensitive data in the smartphone or in the cloud via the smartphone. When an attacker tries to access a smartphone of a legitimate user, the system automatically de-authenticates him. Once SmarterYou decides that the smartphone is now being used by someone other than the legitimate user, the system can perform defensive responses as described earlier. 
Similarly, if the legitimate user is misclassified, several mechanisms for re-instating her are possible, such as two-channel or multi-factor authentication, or requiring an explicit login again, possibly with a biometric, to unlock the system.

\noindent{\bf{Retraining Models}}: The behavioral patterns of SmarterYou users could be changed some time after the initial model training. So it is necessary to retrain users' models to prevent false alarms due to legitimate behavioral drift. SmarterYou provides a model retraining mechanism, which can automatically and continuously retrain the models based on the authentication performance. We define a metric called Confidence Score (CS) to measure if it is necessary to retrain the model. If so, SmarterYou will again upload the legitimate user's latest authentication feature vectors to the cloud server, and update the new models from the training module. It is important to note that adversaries can also exploit this mechanism to retrain the authentication models and achieve accesses to sensitive data with the smartphone. We use multi-factor authentication to prevent these potential vulnerabilities (details in Section~\ref{sec:retrain_model}).

\begin{table}\scriptsize
\centering
\caption{Fisher scores of different sensors.}
\begin{tabular}{|c|c|c|} \hline
       & Smartphone & Smartwatch \\ \hline
Acc(x) & 3.13       & 3.62     \\ \hline
Acc(y) & 0.8        & 0.59\\ \hline
Acc(z) & 0.38       & 0.89\\ \hline
Mag(x) & 0.005      & 0.003\\ \hline
Mag(y) & 0.001      & 0.0049\\ \hline
Mag(z) & 0.0025     & 0.0002\\ \hline
Gyr(x) & 0.57       & 0.24\\ \hline
Gyr(y) & 1.12       & 1.09\\ \hline
Gyr(z) & 4.074      & 0.59\\ \hline
Ori(x) & 0.0049     & 0.0027\\ \hline
Ori(y) & 0.002      & 0.0043\\ \hline
Ori(z) & 0.0033     & 0.0001\\ \hline
Light  & 0.0091     & 0.0428\\ \hline
\end{tabular}
\label{table:fisher_score}
\end{table}
\begin{figure}[!t]
\centering
\includegraphics[width=2.5in,height=1in]{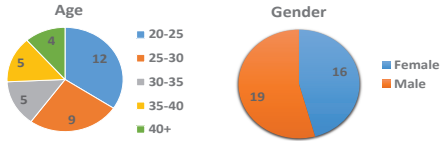}
\DeclareGraphicsExtensions.
\caption{Demographics of the participants}
\label{fig:demographics}
\end{figure}

\subsection{Security Protections}

\hank{
\noindent{\bf{Protecting data in transit}}. \ruby{Since sensitive data are being transmitted between smartwatches, smartphones and cloud servers, secure communications protocols must be used to provide confidentiality and integrity protection against network adversaries.} For instance, an initialization key is exchanged when the smartwatch is paired with the smartphone using Bluetooth. New keys derived from this key can also be used to encrypt and hash the raw data transmitting between smartwatch and smartphone via Bluetooth. The communication channels between smartphones and cloud servers are protected by SSL/TLS protocols.}
 
\hank{\ruby{
\noindent{\bf{Protecting data at rest (i.e., in storage)}}. For data stored in the smartphones or cloud servers, cryptographic encryption and hashing operations are used to prevent the attackers from stealing or modifying data. }}

\hank{
\noindent{\bf{Protecting data and code at runtime.}}
The smartphone and Authentication Server must also provide a secure environment for running the SmarterYou authentication System. Since most smartphones use ARM processors, smartphones can exploit the ARM TrustZone \cite{arm2009security} feature to place the authentication Testing Module in the Secure World and isolate it from other apps in the Normal World. 
Since cloud servers tend to use Intel processors, the trusted Authentication Server can set up secure enclaves by using Intel Sofware Guard eXtensions (SGX) \cite{mckeen2013innovative} for the training and retraining modules for SmarterYou, and for securely accessing and using sensitive behavioral measurements from many smartphone users.}

\begin{figure}[!t] \centering
\subfigure[Smartphone]{
\label{fig:p-value-phone}
\includegraphics[trim =2cm 5cm 1cm 5cm, clip=true, width=0.8\columnwidth ]{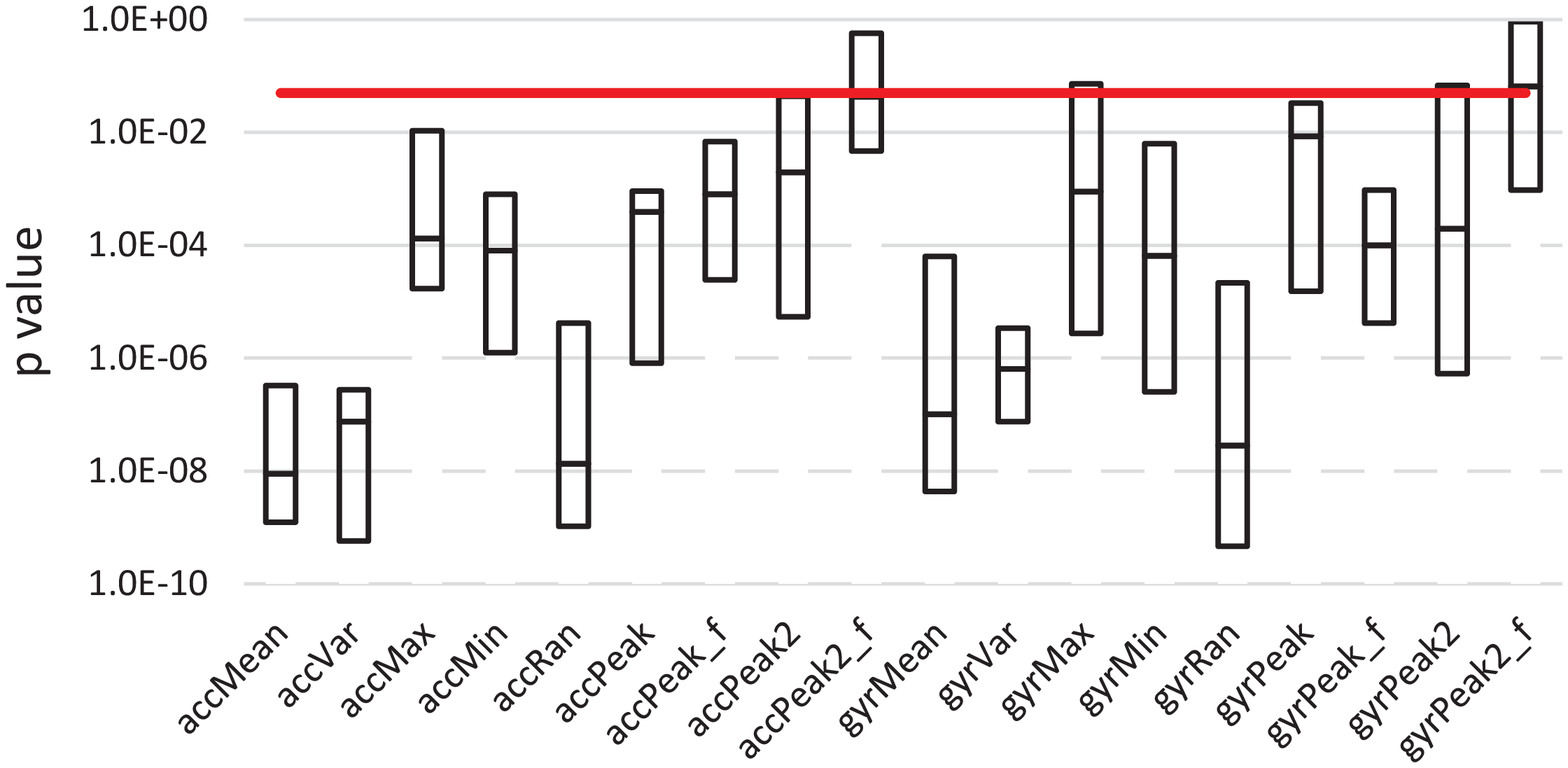}}
\subfigure[Smartwatch]{
\label{fig:p-value-watch}
\includegraphics[trim =2cm 4cm 1cm 4cm, clip=true, width=0.8\columnwidth ]{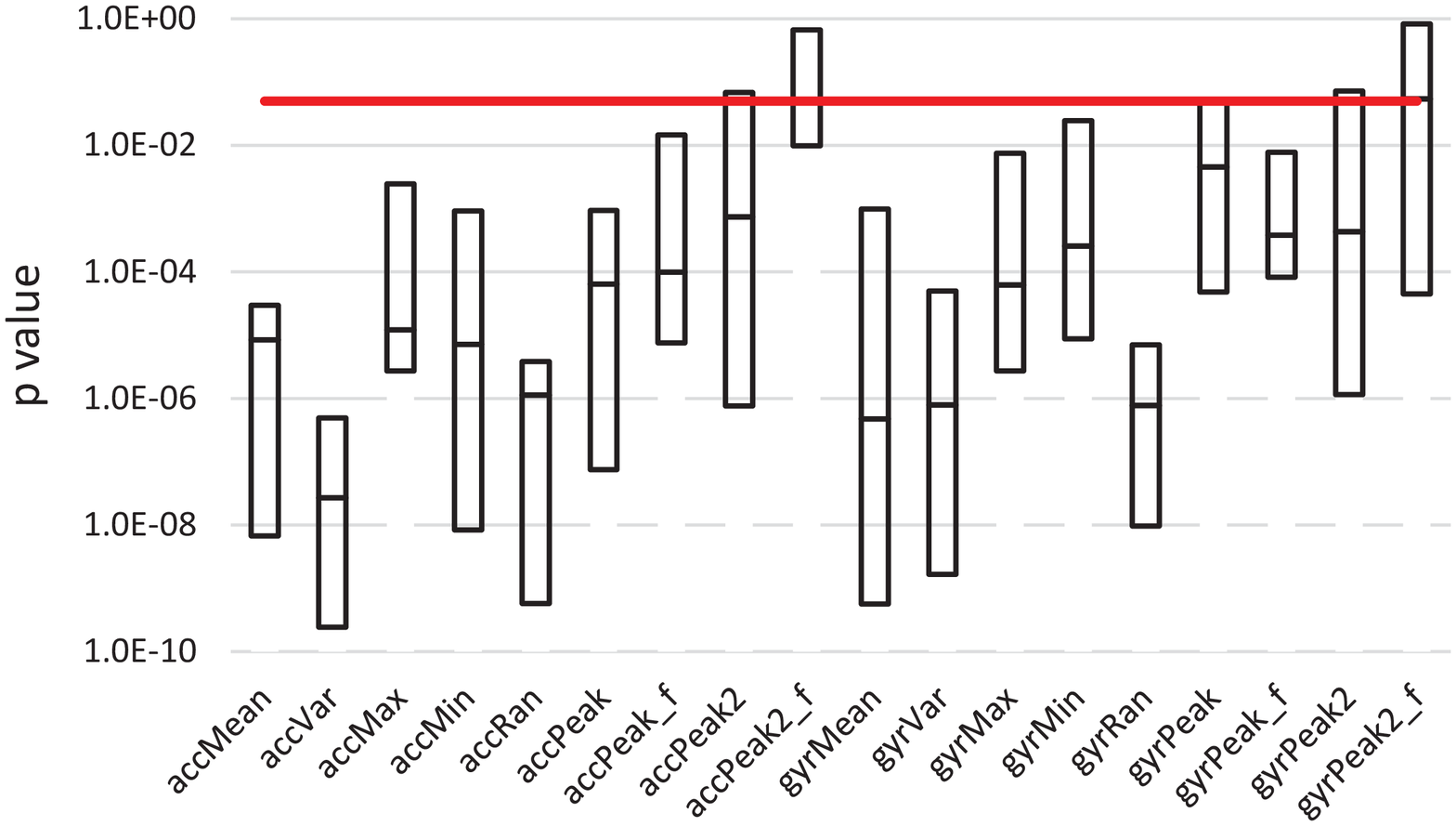}}
\DeclareGraphicsExtensions.
\caption{KS test on sensor features.}
\label{fig:p-value}
\end{figure}

\section{Design Alternatives}\label{sec:experiments}
Although we have outlined the basic architecture for our system, there are many design parameters that have yet to be chosen. Our goal is to get the highest authentication accuracy using the most commonly available sensors and computationally simple algorithms, to facilitate rapid deployment. What sensors should we use?  What features of the raw sensor data streams are best? Can sensors from different devices help improve accuracy? Can contexts improve authentication accuracy, and if so, what are the simplest contexts that give the best accuracy? Which machine learning algorithms are best? Below, we systematically evaluate alternatives for each of these design choices.

\begin{table*}\scriptsize
\centering
\caption{Correlations between each pair of features. The upper triangle is the correlation between features in the smartphone, while the lower triangle is the correlation between features in the smartwatch.}
\begin{tabular}{|c|c|c|c|c|c|c|c|c|c|c|c|c|c|c|c|c|c|} \hline
& & \multicolumn{8}{|c|}{Accelerometer}  & \multicolumn{8}{|c|}{Gyroscope} \\ \hline
& &Mean &Var &Max &Min& Ran& Peak& Peak\_f& Peak2& Mean& Var& Max& Min& Ran& Peak& Peak\_f& Peak2\\\hline
\multirow{8}{*}{\rotatebox{90}{Accelerometer}} & Mean&     &0.39& 0.35& {\bf \textcolor{red}{0.59}}& 0.27& -0.12& -0.15& 0.31& 0.30& 0.17& 0.11& -0.26& 0.13& 0.27& 0.04& 0.34\\ \cline{2-18}
& Var & 0.11&    & 0.28& -0.26& {\bf \textcolor{red}{0.90}}& 0.35& 0.30& 0.41& 0.12& -0.29& 0.10& 0.16& -0.33& 0.25& 0.20&	0.18 \\ \cline{2-18}
& Max&	0.42& 0.37&     & -0.22& {\bf \textcolor{red}{0.78}}& 0.35& 0.23& 0.43& 0.07& 0.32& 0.39& 0.16& 0.19& 0.25& 0.29& 0.23 \\ \cline{2-18}
& Min& 0.31& -0.23& -0.36&    & -0.34& -0.44& -0.43& 0.14& 0.18& -0.10& 0.38& 0.05& -0.32& 0.32& 0.15& 0.05 \\ \cline{2-18}
& Ran& 0.43& {\bf \textcolor{red}{0.94}}& {\bf \textcolor{red}{0.59}}& 0.22&       & 0.28& 0.47& 0.37& 0.22& 0.11& 0.35& 0.21& 0.12& 0.18& -0.08& 0.30 \\ \cline{2-18}
& Peak& -0.02& 0.21& 0.24& -0.33& -0.04&   & 0.19& 0.03& 0.35& 0.23& 0.31& 0.09& 0.30& 0.28& 0.37& 0.21 \\ \cline{2-18}
& Peak\_f& 0.28& -0.04& 0.39& 0.16& 0.43& 0.21&  & 0.09& 0.27& 0.34& 0.10& 0.30& 0.20& 0.05& 0.11&	0.17 \\ \cline{2-18}
& Peak2& -0.16& -0.08& 0.33& 0.44& 0.17& 0.26& -0.32&  & 0.16& 0.31& 0.15& 0.09& 0.29& 0.12& 0.23& 0.28 \\ \hline
\multirow{8}{*}{\rotatebox{90}{Gyroscope}}& Mean& 0.33& 0.31& 0.20& 0.27& 0.27& 0.34& 0.09& 0.16&      & 0.20& 0.31& 0.39& 0.36& 0.21& -0.06& -0.18 \\ \cline{2-18}
& Var& 0.18& 0.35& 0.18& 0.05& 0.35& 0.05& 0.19& 0.12& 0.31&       & -0.15& 0.04& {\bf \textcolor{red}{0.95}}& 0.08& 0.34& -0.51 \\ \cline{2-18}
& Max& 0.32& 0.34& 0.36& 0.30& -0.17& 0.33& 0.22& 0.17& -0.27& 0.21&       & 0.37& {\bf \textcolor{red}{0.68}}& 0.42& -0.27& 0.38 \\ \cline{2-18}
& Min& 0.32& 0.16& 0.18& 0.04& 0.29& 0.24& 0.12& 0.13& -0.14& 0.47& 0.37&       & 0.34& 0.18& 0.15& -0.34 \\ \cline{2-18}
& Ran& 0.13& 0.04& -0.19& 0.11& 0.29& 0.09& -0.24& 0.19& 0.03&{\bf \textcolor{red}{0.89}}&{\bf \textcolor{red}{0.60}}& -0.14&       & -0.21& 0.12& 0.23 \\ \cline{2-18}
& Peak& 0.07& 0.15& -0.33& 0.18& 0.35& 0.06& 0.33& 0.30& 0.25& -0.18& 0.41& 0.02& -0.38&      & 0.33& 0.16\\ \cline{2-18}
& Peak\_f& 0.21& 0.17& 0.23& 0.26& -0.13& 0.13& 0.36& 0.16& 0.32& 0.36& -0.20& 0.32& 0.24& 0.18&    & 0.07 \\ \cline{2-18}
& Peak2& 0.33& 0.07& 0.30& 0.16& 0.32& 0.21& 0.35& 0.15& -0.29& 0.12& -0.10& 0.39& 0.34& 0.12& -0.19& \\ \hline
\end{tabular}
\label{table:correlation_ppww}
\end{table*}

\subsection{Experimental settings} \label{sec:setting}
We perform different types of experiments with $35$ \new{participants}, using Nexus 5 smartphones and Moto 360 smartwatches. \new{We recorded the demographics (gender, and age range) of the participants and show them in Figure~\ref{fig:demographics}.} We collected sensor data from different sensors in the smartphone and the smartwatch, with a sampling rate of $50$ Hz. The different types of experiments (free-form usage, lab experiments and attacker usage) will be discussed in detail in the sub-sections they are used, as we attempt to answer the above questions on the design parameters of our implicit authentication system. All experimental results in the following sub-sections are based on the free-form use of the smartphone and smartwatch for two weeks, except the experiments for context detection (where lab conditions are used) and the masquerading attacks (where attacker usage is imitated). Free-form usage means the users can use the devices with no restrictions, as they normally would in their daily lives.

In our collected data for the machine learning algorithms, we used 10-fold cross-validation to generate the training data and testing data sets for evaluating the authentication performance, i.e., $9/10$ data would be used as the training data and the remaining $1/10$ is used as the testing data. To extensively investigate the performance of our system, we repeated such cross-validating mechanisms for $1000$ iterations and averaged the experimental results.

We also discuss the complexity of our system and the impact on the battery drainage (Section \ref{sec:overhead}). Finally, we discuss re-training authentication models (Section \ref{sec:retrain_model}) due to users' behavioral drift.

\subsection{Which sensors to use?} \label{sec:sensor}
Mobile sensing technology has matured to a state where collecting many measurements through sensors in smartphones is now becoming quite easy through, for example, Android sensor APIs. Mobile sensing applications, such as the CMU MobiSens\cite{cc1}, run as a service in the background and can constantly collect sensors' information from smartphones. Sensors can be either hard sensors (e.g., accelerometers) that are physically-sensing devices, or soft sensors that record information of a phone's running status (e.g., screen on/off). Thus, practical sensors-based user authentication can be achieved today. But which sensors should we select?

We use Fisher scores (FS)\cite{duda2012pattern} to help select the most promising sensors for user authentication. FS is one of the most widely used supervised feature selection methods due to its excellent performance. 
The Fisher Score enables finding a subset of features, such that in the data space spanned by the selected features, the distances between data points in different classes are as large as possible, while the distances between data points in the same class are as small as possible.
Table \ref{table:fisher_score} shows the FS for different sensors that are widely implemented in smartphones and smartwatches. We found that the magnetometer, orientation sensor and light sensor have lower FS because they are influenced by the environment. This can introduce various background noise unrelated to the user's behavioral characteristics, e.g., the magnetometer may be influenced by magnets. Therefore, we select two sensors, the accelerometer and gyroscope, because they have higher FS and furthermore, are the most common sensors built into current smartphones and smartwatches~\cite{google}.

These two sensors also represent different information about the user's behavior: 1) the accelerometer records coarse-grained motion patterns of a user, such as how she walks \cite{cc7}; and 2) the gyroscope records fine-grained motions of a user such as how she holds a smartphone \cite{cc8}. Furthermore, these sensors do not need the user's permissions, making them useful for continuous background monitoring in implicit authentication scenarios.

\begin{table*}\scriptsize
\centering
\caption{Correlations between smartphone and smartwatch. Row labels are the features from smartwatch and column labels are the features from smartphone.}
\begin{tabular}{|c|c|c|c|c|c|c|c|c|c|c|c|c|c|c|c|} \hline
& & \multicolumn{7}{|c|}{Smartphone Accelerometer}  & \multicolumn{7}{|c|}{Smartphone Gyroscope} \\ \hline
& &Mean &Var &Max &Min& Peak& Peak\_f& Peak2& Mean& Var& Max& Min& Peak& Peak\_f& Peak2\\\hline
\multirow{7}{*}{\rotatebox{90}{\tabincell{c}{Smartwatch \\Accelerometer}}}& Mean& 0.08& 0.33& -0.23& 0.20& 0.26& 0.10& 0.42& 0.27& -0.31& -0.10& 0.03& 0.13& -0.19& 0.06 \\ \cline{2-16}
& Var & -0.29& 0.23& 0.09& -0.08& -0.21& 0.27& -0.24& 0.04& 0.39& 0.26& 0.05& 0.17& 0.15& 0.37\\ \cline{2-16}
& Max&	0.35& -0.05& -0.02& -0.34& -0.15& -0.33& 0.20& -0.25& 0.24& 0.09& 0.26& -0.32& 0.23& -0.22\\ \cline{2-16}
& Min& -0.24& 0.29& -0.34& 0.21& -0.37& 0.39& 0.05& 0.30& 0.04& -0.33& -0.32& -0.15& -0.23& -0.13 \\ \cline{2-16}
& Peak& -0.08& -0.11& 0.40& 0.08& -0.07& -0.33& -0.35& -0.17& 0.21& 0.24& -0.29& 0.08& -0.28& 0.21 \\ \cline{2-16}
& Peak\_f& 0.11& -0.21& 0.03& -0.10& 0.33& 0.07& 0.34& -0.22& -0.18& 0.04& 0.32& -0.07& -0.12& -0.31\\ \cline{2-16}
& Peak2& -0.26& -0.16& -0.08& 0.14& -0.32& -0.26& 0.24& 0.24& -0.24& 0.41& 0.15& -0.37& -0.12& -0.32 \\ \hline
\multirow{7}{*}{\rotatebox{90}{\tabincell{c}{Smartwatch \\Gyroscope}}}& Mean& 0.02& 0.13& -0.16& 0.08& 0.36& 0.37& -0.26& -0.31& 0.20& -0.31& 0.33& 0.37& -0.24& 0.26 \\ \cline{2-16}
& Var& 0.16& 0.29& -0.33& -0.26& 0.03& -0.30& -0.10& -0.26& 0.03& 0.05& 0.02& -0.29& 0.27& 0.21 \\ \cline{2-16}
& Max& -0.12& -0.30& 0.22& 0.21& -0.14& -0.20& -0.03& 0.10& 0.12& 0.05& 0.31& 0.30& 0.32& -0.28 \\ \cline{2-16}
& Min& 0.07& -0.22& -0.18& 0.19& -0.29& 0.30& 0.11& 0.15& 0.06& 0.29& -0.33& -0.11& -0.04& -0.13\\ \cline{2-16}
& Peak& 0.28& -0.21& -0.27& 0.34& 0.37& 0.16& 0.23& 0.29& 0.20& 0.04& 0.14& 0.19& -0.10& -0.05\\ \cline{2-16}
& Peak\_f& -0.23& -0.06& -0.25& 0.29& 0.33& 0.18& 0.28& -0.16& 0.25& -0.32& 0.20& -0.04& -0.06& 0.12 \\ \cline{2-16}
& Peak2& 0.13& -0.07& 0.21& -0.27& 0.37& 0.32& -0.11& 0.38& -0.12& -0.22& 0.06& 0.04& 0.33& 0.11 \\ \hline
\end{tabular}
\label{table:correlation_pw}
\end{table*}

\subsection{What sensor features are best?} \label{sec:featrure_selection}
Using the raw sensor data streams from the selected sensors may not be as good as using statistical features derived from these raw sensor data streams. Hence, we segment the sensor data streams into a series of time windows, and compute statistics from both the time domain and the frequency domain for the sensor data values in a time window.
The magnitude of sensor $i$'s data stream in the $k$-th window is denoted $S_i(k)$. For example, the magnitude of an accelerometer data sample $(t,x,y,z)$  is computed as $m=\sqrt{x^2+y^2+z^2}$. 
We implement the Discrete \emph{Fourier} transform (DFT) \cite{boashash2003time} to obtain the frequency domain information.
The frequency domain information is useful and is widely used in signal processing and data analysis, e.g., speech signals and images.

We compute the following statistical features derived from each of the raw sensor streams, in each time window:
\begin{enumerate}[$\bullet$]
\item Mean: Average value of the sensor stream
\item Var: Variance of the sensor stream
\item Max: Maximum value of the sensor stream
\item Min: Minimum value of the sensor stream
\item Ran: Range of the sensor stream
\item Peak: The amplitude of the main frequency of the sensor stream
\item Peak\_f: The main frequency of the sensor stream
\item Peak2: The amplitude of the secondary frequency of the sensor stream
\item Peak2\_f: The secondary frequency of the sensor stream
\end{enumerate}

We then test the performance of each feature and drop ``bad" features. If a feature can be used to easily distinguish two users, we say the feature is a good feature. For a feature to distinguish two different persons, it is necessary for the two underlying distributions to be different. Hence, for each feature, we test whether this feature derived from different users is from the same distribution. If most pairs of them are from the same distribution, the feature is ``bad" in distinguishing two persons and we drop it.

We use the Kolmogorov-Smirnov test (KS test) \cite{daniel1990applied} to test if two data sets are significantly different. The KS test is a nonparametric statistical hypothesis test based on the maximum distance between the empirical cumulative distribution functions of the two data sets. The two hypotheses of a KS test are:

\textit{$H_0$: the two data sets are from the same distribution}

\textit{$H_1$: the two data sets are from different distributions.}

A KS test reports a $p$-value, i.e. the probability that obtaining the maximum distance is at least as large as the observed one when $H_0$ is assumed to be true. i.e., $H_0$  is accepted. If this $p$-value is smaller than $\alpha$, usually set to $0.05$, we will reject the $H_0$ hypothesis because events with small probabilities rarely happen (rejecting $H_0$ and accepting $H_1$  indicates a ``good" feature for distinguishing users). For each feature, we calculate the $p$-value 
for data points for each pair of users and drop a feature if most of its $p$-values are higher than $\alpha$.

Figure \ref{fig:p-value} shows the testing results for the features in both the smartphone and smartwatch. For each feature, the resulting $p$-values are drawn in a box plot. The
bottom and the top lines of the box denote the lower quartile $Q_1$ and upper quartile $Q_2$, defined as the $25$th and the $75$th percentiles of the $p$-values. The middle bar denotes the median of the $p$-values. The $y$-axes in Figure \ref{fig:p-value} is in logarithmic scale. The red horizontal lines represent the significance level $\alpha = 0.05$. The better a feature is, the more of its box plot is below the red line. It denotes that more pairs are significantly different. From Figure \ref{fig:p-value}, we find that the accPeak2\_f and gyrPeak2\_f in both the smartphone and the smartwatch are ``bad" features, so we drop them. 

Next, we try to drop redundant features, by computing the correlation between each pair of features. A strong correlation between a pair of features indicates that they are similar in describing a user's behavior pattern, so one of the features can be dropped. A weak correlation implies that the selected features reflect different behaviors of the user, so both features should be kept. 

We calculated the Pearson's correlation coefficient between any pair of features. Then, for every pair of features, we took the average of all resulting correlation coefficients over all the users.
Table \ref{table:correlation_ppww} shows the resulting average correlation coefficients. The upper right triangle is the correlation between features in the smartphone, while the lower left triangle is the correlation between features in the smartwatch. We observe that Ran has very high correlation with Var in each sensor on both the smartphone and smartwatch. It means that Ran and Var have information redundancy. Also Ran has relatively high correlation with Max. Therefore, we drop Ran from our feature set.

\subsection{Do multiple devices help?} \label{sec:multiple_devices}

We also study if using data from the same type of sensors (accelerometer and gyroscope), but from different devices is helpful for improving user authentication.  Towards this end, we calculate the correlations between smartphone and smartwatch sensor data in Table \ref{table:correlation_pw}. Since these features do not have strong correlation with each other, it implies that these same sensors on the two devices measure different aspects of a user's behavior, so we keep all these features.  

Hence our feature vector for sensor $i$, in a given time window $k$, for the smartphone, $SP$, is

\begin{equation}
SP_{i}(k)=[SP_{i}^t(k), SP_{i}^f(k)]
\end{equation}
where \hank{$t$ represents the \ruby{time domain, $f$ represents the frequency domain}, and}
\begin{equation}
\begin{aligned}
SP_{i}^t(k)&=[mean(S_{i}(k)),var(S_{i}(k)),max(S_{i}(k)),min(S_{i}(k))] \\ 
SP_{i}^f(k)&=[peak(S_{i}(k)),freq(S_{i}(k)),peak2(S_{i}(k))]
\end{aligned}
\end{equation}
Therefore the feature vector for the smartphone is
\begin{equation}\label{eq:feature}
SP(k)=[SP_{accerometer}(k), SP_{gyroscope}(k)]
\end{equation}
Similarly, we have the the feature vector for the sensor data from the smartwatch, denoted $SW(k)$. Therefore, the authentication feature vector is
\begin{equation}\label{eq:authenticatevector}
Authenticate(k)=[SP(k), SW(k)]
\end{equation}


\subsection{Can Context Detection help?} \label{sec:contextDetection}
Since it seems intuitive that sensor measurements of motion may be different under different contexts, we now consider the minimum contexts that can improve the accuracy of user authentication. To be viable, we need very fast, user-agnostic context detection, since this must now precede user authentication, and we also want to keep real-time computation to an acceptable level. Hence, we try using the same feature vector in Eq. \ref{eq:feature} for the smartphone only (no smartwatch) context detection.
During the user enrollment phase, we feed these feature vectors from all users into the context detection model to train it. During the testing phase, we use this user-agnostic context detection model to detect the current user context.

\subsubsection{Random Forest for context detection}
We experimented with several machine learning algorithms for context detection, and chose the Random forest algorithm \cite{ho1995random}.
This is commonly used in data mining. It creates a model that predicts the value of a target variable based on several input variables. 

\begin{table}\scriptsize
\centering
\caption{Confusion matrix of context detection results using two smartphone sensors.}
\begin{tabular}{|c|c|c|} \hline
\multicolumn{1}{|c|}{Confusion Matrix} & \multicolumn{1}{|c|}{Stationary} & \multicolumn{1}{|c|}{Moving}\\ \hline
Stationary & $99.1\%$ & $0.9\%$ \\ \hline
Moving & $0.6\%$ & $99.4\%$ \\ \hline
\end{tabular}
\label{table:cm_context}
\end{table}

Initially, we tried using four contexts: (1) The user uses the smartphone without moving around, 
e.g., while standing or sitting; (2) The user uses the smartphone while moving. No constraints are set for how the user moves; (3) The smartphone is stationary (e.g., on a table) while the user uses it; (4) The user uses the smartphone on a moving vehicle, e.g., train. However, we found that these four contexts can not be easily differentiated: contexts (3) and (4) are easily misclassified as context (1), since (1), (3) and (4) are all relatively stationary \hank{(e.g., when moving at a stable speed)}, compared to context (2). Therefore, we combined contexts (1), (3) and (4) into one stationary context, and left (2) as the moving context. The resulting confusion matrix in Table \ref{table:cm_context} showed a very high context detection accuracy of over 99\% with these 2 simple contexts. The context detection time was also very short - less than 3 milliseconds.  

For these context training and testing experiments, we had users use their smartphones in fixed contexts under controlled lab conditions. 
Users were asked to use the smartphone and the smartwatch freely under each context for 20 minutes. They were told to stay in the current context until the experiment is finished. \new{Note that such recording process is only needed for developing the context detection model and is not required for normal use in real-world scenarios.}
We use these data  from the different users to train the context detection model in a user-agnostic manner. That is, when we perform context detection for a given user, we use a context detection model (i.e., classifier) that was trained with other users' data. This enables us to detect the context of the current user prior to authenticating her. 
For the Random Forest algorithm, we use 10-fold cross-validation to get the results in Table \ref{table:cm_context}.

\begin{table}\scriptsize
\centering
\caption{Authentication performance with different Machine Learning algorithms.}
\begin{tabular}{|c|c|c|c|} \hline
Method & FRR  & FAR  & Accuracy\\ \hline
KRR    &  $0.9\%$  & $2.8 \%$	&  $98.1\%$ \\ \hline
SVM    &  $2.7\%$  & $2.5 \%$	&  $97.4\%$ \\ \hline
Linear Regression    &  $12.7\%$  & $14.6 \%$	&  $86.3\%$ \\ \hline
Naive Bayes    &  $10.8\%$  & $13.9 \%$	&  $87.6\%$ \\ \hline
\end{tabular}
\label{table:ml_compare}
\end{table}

\begin{figure}[!t] \centering
\subfigure[Stationary]{
\label{S5_FNR} 
\includegraphics[width=1.5in,height=0.9in]{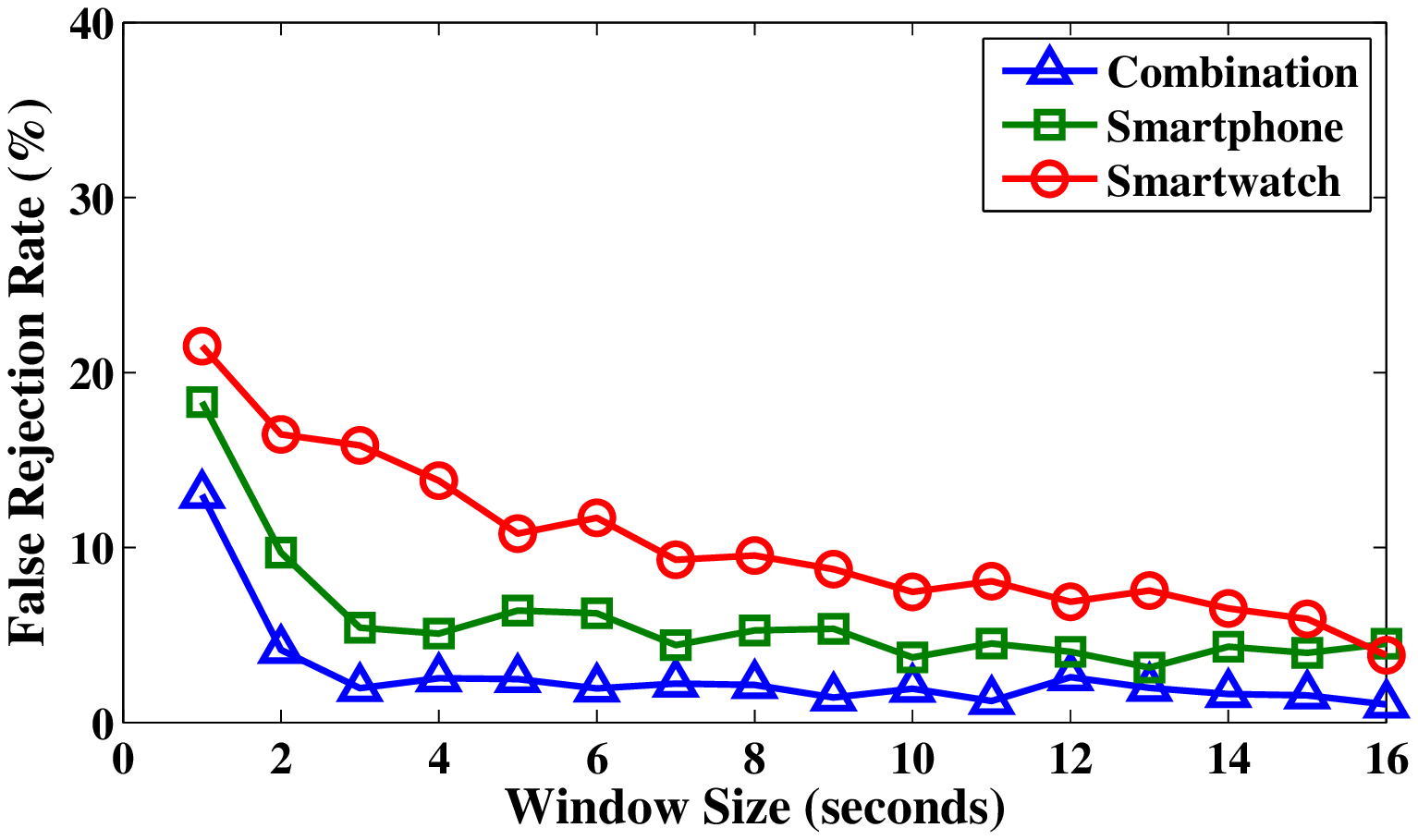}}
\subfigure[Moving]{
\label{S4_FNR} 
\includegraphics[width=1.5in,height=0.9in]{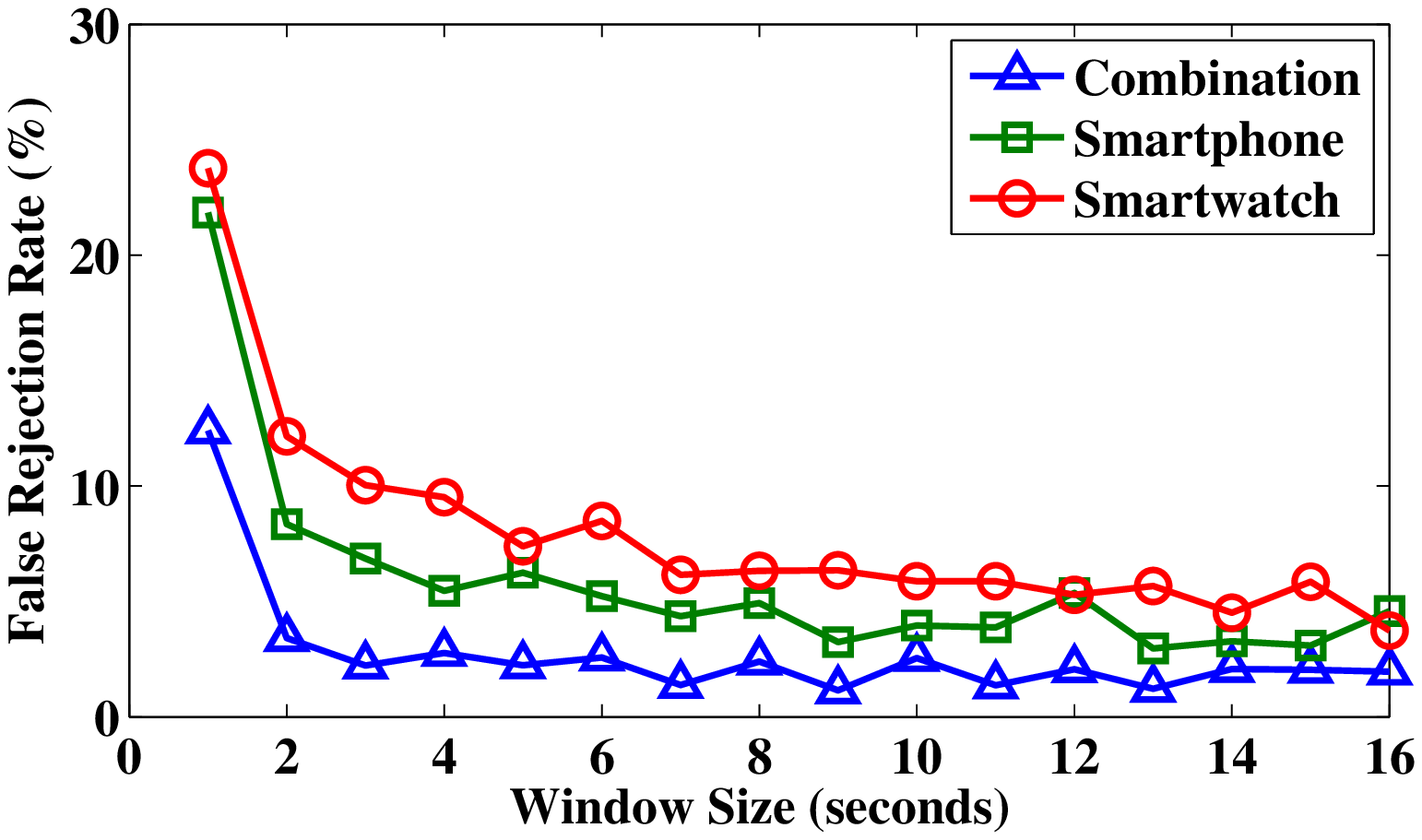}}
\subfigure[Stationary]{
\label{S5_FPR} 
\includegraphics[width=1.5in,height=0.9in]{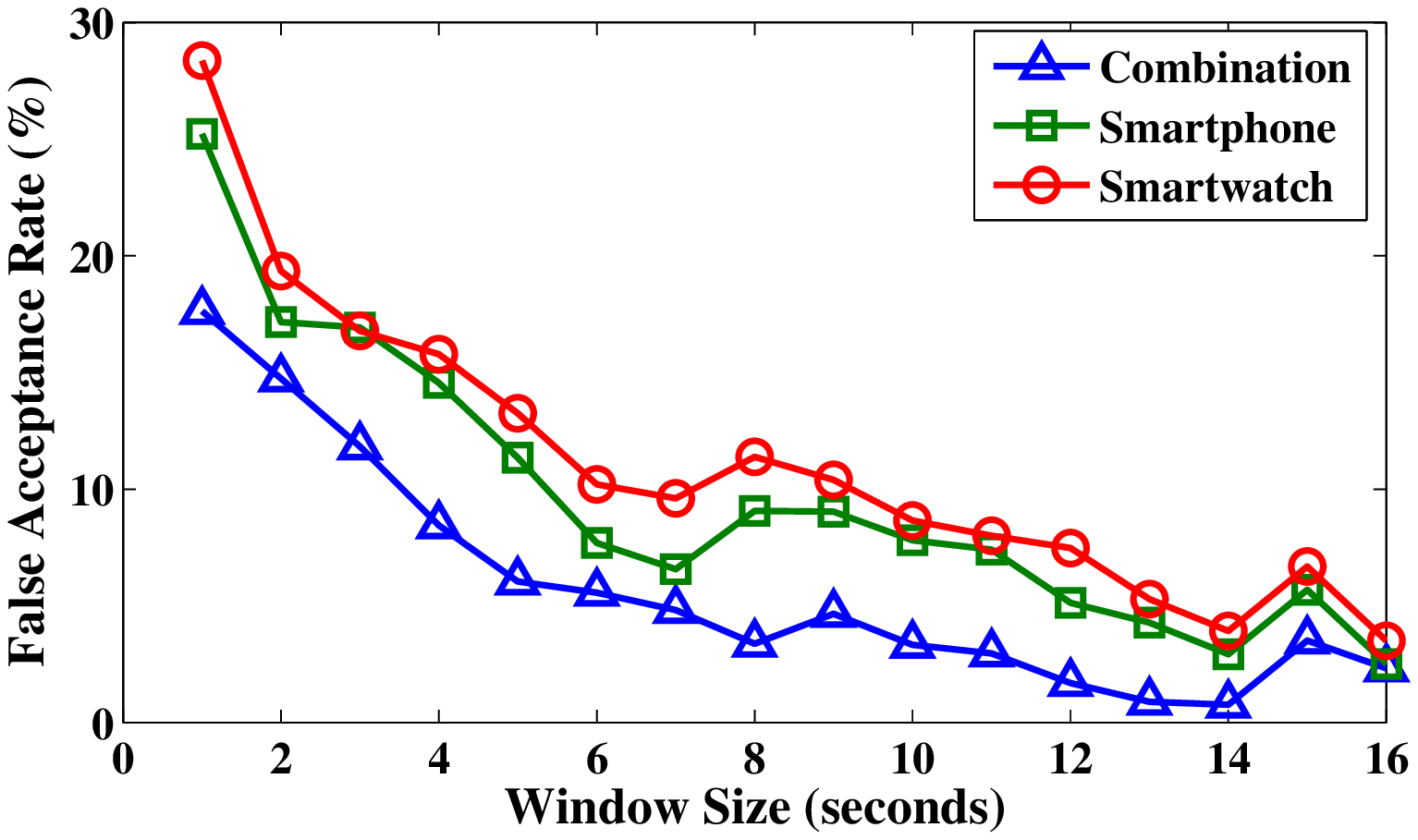}}
\subfigure[Moving]{
\label{S4_FPR} 
\includegraphics[width=1.5in,height=0.9in]{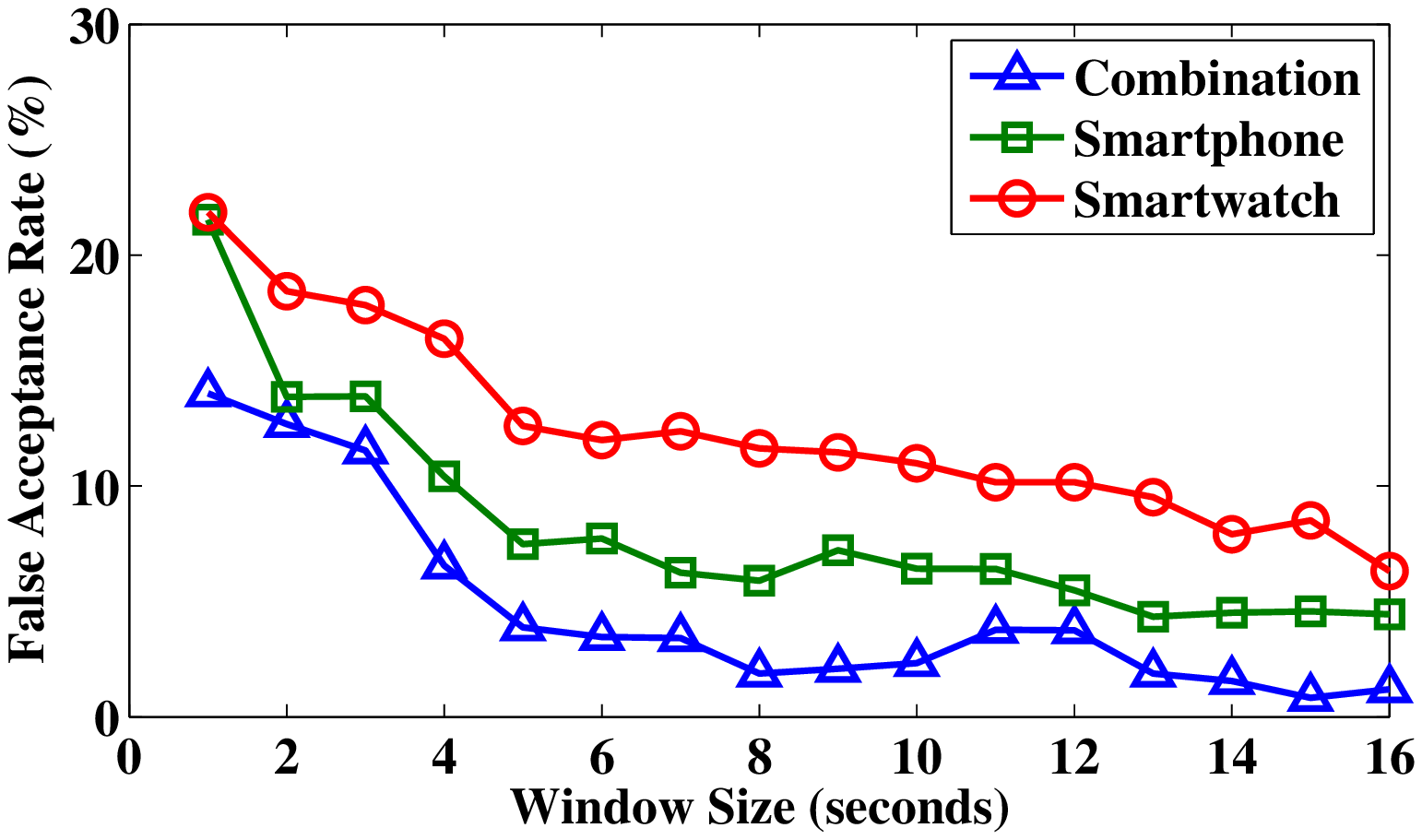}}
\caption{FRR and FAR with different window sizes under two contexts. (a) and (b) are the FRRs under different contexts. (c) and (d) are the FARs under different contexts. Both the FRR and FAR become stable when the window size is larger than $6$ seconds.}
\label{fig:window_size}
\end{figure}

\subsection{User Authentication Algorithms} \label{sec:authentication}
\subsubsection{Features}

We now ask whether such simple, fast and user-agnostic contexts (stationary versus moving) can significantly  improve the accuracy of user authentication? If so, to what extent?
For this, we did different experiments, where the users could use their smartphones and smartwatches as they normally do in their daily lives, without any constraints on the contexts under which they used their devices.
Users were invited to take our smartphone and smartwatch for one to two weeks, and use them under free-form, real-use conditions.

We evaluate the accuracy of user authentication when only the smartphone's sensor features from the accelerometer and gyroscope were used, and when both the smartphone and smartwatch's sensor features were used.  The former had feature vectors with $7\times 2=14$ elements, while the latter had feature vectors with $7\times 2 \times 2=28$ elements. 

\subsubsection{Kernel Ridge Regression algorithm}\label{sec:krr}
Here we tried different machine learning algorithms, and found the Kernel Ridge Regression (KRR) machine learning algorithm to give the best results.
Table \ref{table:ml_compare} shows user authentication results for a sample of state-of-the-art machine learning techniques: KRR, Support Vector Machines (SVM), linear regression,  and naive Bayes. We see that KRR achieves the best accuracy. SVM also achieves high accuracy but the computational complexity is much higher than KRR (shown in Section \ref{sec:overhead}). Linear regression and naive Bayes have significantly lower accuracy 
compared to KRR and SVM.

Kernel ridge regressions (KRR) have been widely used for classification analysis \cite{suykens2002krr, an2007face,lee2015implicit,lee2016implicit}. 
The advantage of KRR is that the computational complexity is much less than other machine learning methods, e.g., SVM. The goal of KRR is to learn a model that assigns the correct label to an unseen testing sample. This can be thought of as learning a function $f: X \rightarrow Y$ which maps each data $x$ to a label $y$. The optimal classifier can be obtained analytically according to
\begin{equation}\label{eq2}
\bm{w}^{*}=\mathrm{argmin}_{\bm{w}\in\mathbb{R}^d}\rho\|\bm{w}\|^2+\sum_{k=1}^N (\bm{w}^T {\bm{x}}_k -y_k)^2
\end{equation}
where $N$ is the data size and $\bm{x}^{M\times1}_k$ represents the transpose of $Authenticate(k)$, 
the authentication feature vector,
and $M$ is the dimension of the authentication feature vector. Let $\bm{X}$ denote a $M\times N$ training data matrix $\bm{X}= [{\bm{x}}_1,{\bm{x}}_2, \cdots, {\bm{x}}_N]$. Let $\bm{y}= [{\bm{y}}_1,{\bm{y}}_2, \cdots, {\bm{y}}_N]$.
$\bm{\vec{\phi}(\bm{x_i})}$ denotes the kernel function, which maps the original data $\bm{x_i}$ into a higher-dimensional ($J$) space. In addition, we define $\bm{\Phi} = [\bm{\vec{\phi}(\bm{x_1})}\bm{\vec{\phi}(\bm{x_2})}\cdots \bm{\vec{\phi}(\bm{x_N})}] $ and $\bm{K}=\bm{\Phi}^T\bm{\Phi}$.
This objective function in Eq.~\ref{eq2} has an analytic optimal solution \cite{suykens2002krr} where
\begin{equation}\label{optimal}
\bm{w}^{*} =\bm{\Phi}[\bm{K}+\rho \bm{I_N}]^{-1} \bm{y}
\end{equation}
By utilizing certain matrix transformation properties, the computational complexity for computing the optimal $\bm{w}^{*}$ in Eq.~\ref{optimal} can be largely reduced from $O(N^{2.373})$ to $O(M^{2.373})$, which we will carefully discuss in Section~\ref{sec:overhead}.  This is a huge reduction since N=800 data points in our experiments, and M = 28 features in our authentication feature vector.

\begin{figure}[!t] \centering
\subfigure[Stationary]{
\label{S5_Acc} 
\includegraphics[width=1.5in,height=0.9in]{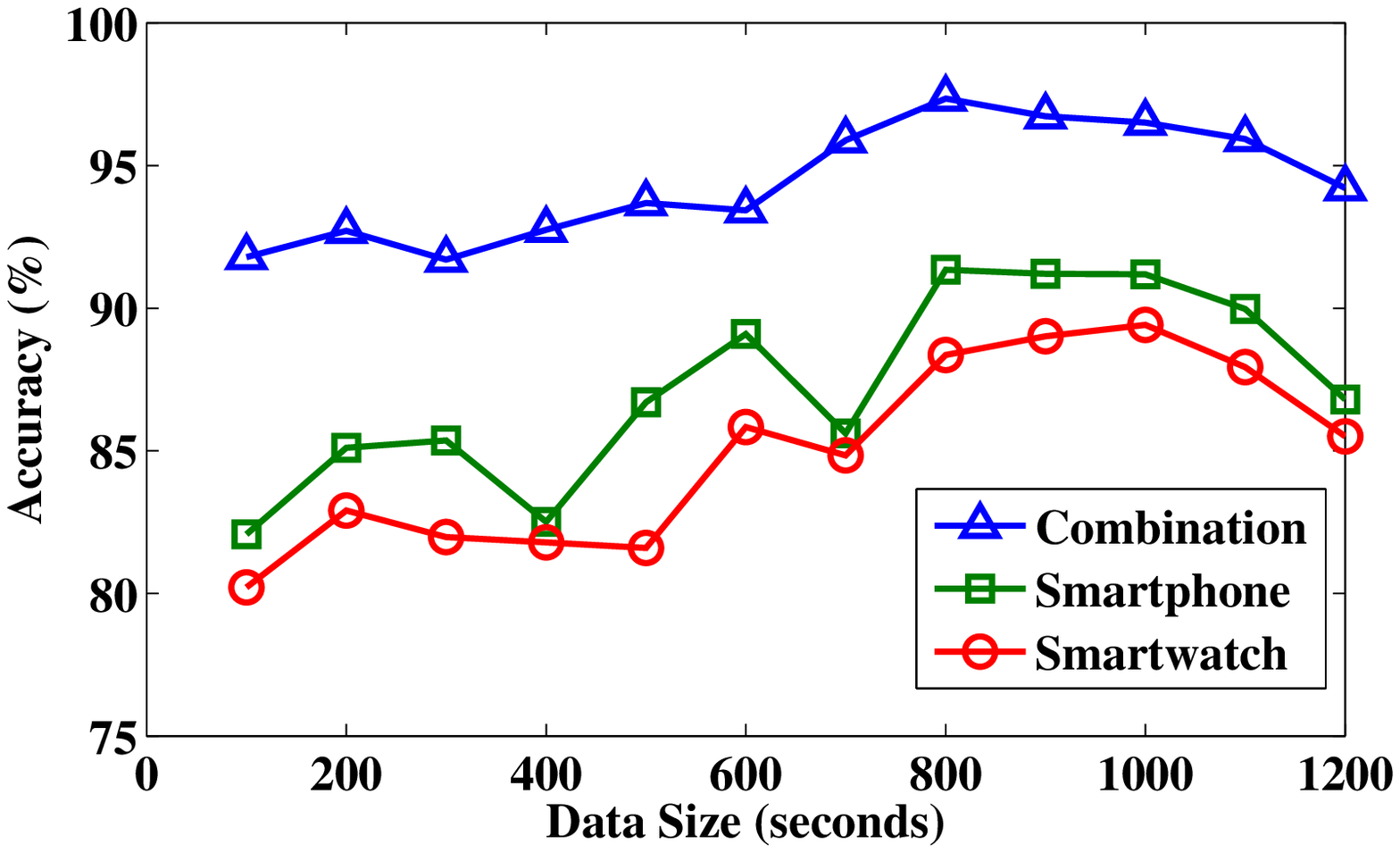}}
\hspace{-1.5em}
\subfigure[Moving]{
\label{S4_Acc} 
\includegraphics[width=1.5in,height=0.9in]{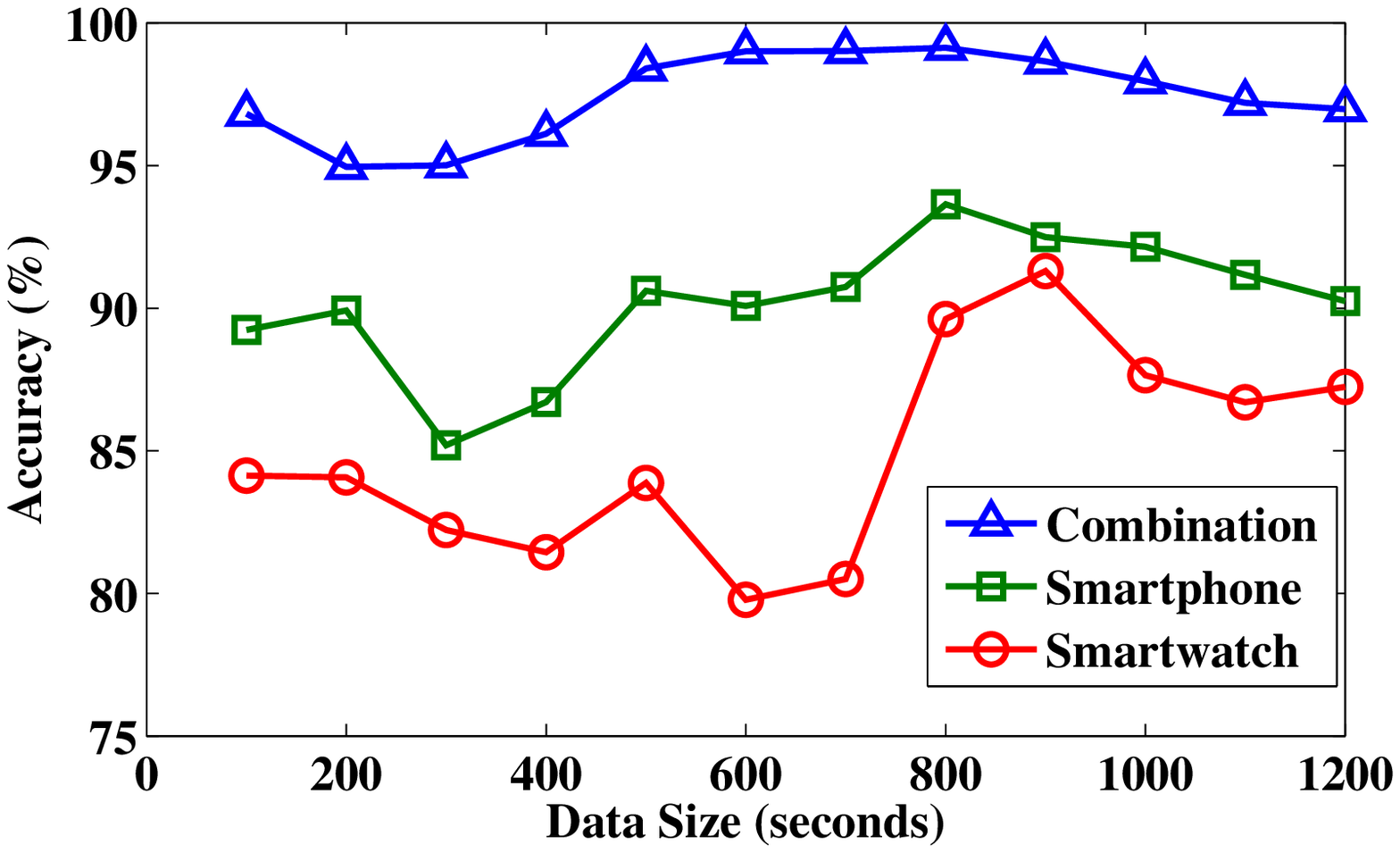}}
\caption{Accuracy with different data sizes under the two contexts. We observe that the best accuracy happens when the data size is around $800$. The accuracy decreases after the training set size is larger than $800$ because a large training data set is likely to cause over-fitting in the machine learning algorithms.}
\label{fig:data_size} 
\end{figure}

\subsubsection{System Parameters}\label{sec:parameter}
We need to decide on two important parameters in the system, the window size and the size of the dataset. We empirically derive the ``optimal" values for these parameters.

\noindent{\bf Window Size.\label{windowsize}}

The window size is an important system parameter, which determines the time that our system needs to perform an authentication, i.e., window size directly determines our system's authentication frequency.

For each context, we vary the window size from 1 second to 16 seconds. \hank{Given a window size and a detected context, for each target user, we utilize 10-fold cross-validation for training and testing.} Here, we utilize the false reject rate (FRR) and false accept rate (FAR) as metrics to evaluate the authentication accuracy of our system. FRR is the fraction of the legitimate user's data that are misclassified as other users' data. FAR is the fraction of other users' data that are misclassified as the legitimate user's. For security protection, a large FAR is more harmful than a large FRR. However, a large FRR would degrade the usage convenience. Therefore, we investigate the influence of the window size on FRR and FAR, in choosing a proper window size.

Figure~\ref{fig:window_size} shows that the FRR and FAR for each context become stable when the window size is greater than 6 seconds. The smartphone has better (lower) FRR and FAR than the smartwatch. The combination of the smartphone and smartwatch has the lowest FRR and FAR, and achieves the best authentication performance than using each alone.

\noindent{\bf Data Size.\label{datasize}}

Another important system parameter is the size of the data set, which also affects the overall authentication accuracy because a larger training data set provides the system more information. According to our observations above, we set the window size as $6$ seconds. We ranged the training set sizes, from $100$ to $1200$ and show the experimental results in Figure~\ref{fig:data_size}. 
We see that as the training set size increases, the accuracy first increases, approaching a maximum accuracy point, and then decreases. The maximum accuracy happens when the data size is around $800$. The accuracy decreases after the training set size is larger than $800$ because a large training data set is likely to cause over-fitting in the machine learning algorithms so that the constructed training model would introduce more errors than expected. Comparing the three lines in each figure, we also find that using more devices provides extra information that improves authentication accuracy.

\begin{table}\scriptsize
\centering
\caption{The FRR,FAR and accuracy under two contexts with different devices.}
\begin{tabular}{|c|c|c|c|c|} \hline
Context& Device 	 & FRR  & FAR  & Accuracy\\ \hline
w/o context & Smartphone  &  $15.4\%$ & $17.4\%$	&  $83.6\%$ \\ \cline{2-5}
        	& Combination &  $7.3\%$  & $9.3 \%$	&  $91.7\%$ \\ \hline
w/ context   & Smartphone  &  $5.1\%$ & $8.3 \%$	    &  $93.3\%$ \\ \cline{2-5}
	        & Combination &  $0.9\%$  & $2.8 \%$	&  $98.1\%$ \\ \hline
\end{tabular}
\label{table:krr}
\end{table}
\subsubsection{User Authentication Evaluation with KRR}
We now show the overall authentication performance of our system in Table~\ref{table:krr} by setting the window size as $6$ seconds and the data size as $800$ (from Section~\ref{sec:parameter} results).

From Table~\ref{table:krr}, we have the following interesting observations: 
(1) {\bf{SmarterYou works well with just the smartphone, even without contexts}}: by using only the smartphone without considering any context, our system can achieve authentication accuracy up to $83.6\%$.
(2) {\bf{Auxiliary devices are helpful}}: by combining sensor data from the smartwatch with the smartphone sensor data, the authentication performance increases significantly over that of the smartphone alone, reaching $91.7\%$ accuracy, with better FRR and FAR.
(3) {\bf{Context detection is beneficial for authentication}}: the authentication accuracy is further improved, when we take the finer-grained context differences into consideration, reaching $93.3\%$ accuracy with the smartphone alone, and $98.1\%$ accuracy with the combination of smartphone and smartwatch data.

We also found that the overall time for implementing 
context detection followed by user authentication is less than $21$ milliseconds. 
This is a fast user authentication testing time, with excellent authentication accuracy of 98\%, making our system efficient and applicable in real world scenarios.

\subsection{Masquerading attacks}\label{sec:security}
Our third set of experiments was designed to analyze our system's performance in defending against some real world attacks (e.g., masquerading or mimicry attacks). \new{We consider the worst case situation where we assume the attacker is able to monitor and record the victim's behavior. Thus the attacker can try his best to learn the victim's behavior.} In these experiments, we asked each subject to be a malicious adversary whose goal was to mimic the victim user's behavior to the best of his/her ability. One user's data was recorded and his/her model was built as the legitimate user. The other users tried to mimic the legitimate user and cheat the system to let them be authenticated as the victim user. The victim user was recorded by a VCR. Subjects were asked to watch the video and mimic the behavior. Both the adversary and the legitimate user performed the same tasks, and the user's behavior is clearly visible to the adversary. Such an attack is repeated $20$ times for each legitimate user and his/her `adversaries'.

Recall that the goal of an attacker is to get access to the sensitive information stored in the smartphone, or in the cloud accessed through the smartphone. As we have shown in Figure \ref{fig:window_size} and Table \ref{table:krr}, SmarterYou achieves very low FARs when attackers attempt to use the smartphone with their own behavioral patterns.

Now, we show that SmarterYou is even secure against the masquerading attacks where an adversary tries to mimic the user's behavior. Here, \textit{`secure'} means that the attacker cannot cheat the system via performing these spoofing attacks and the system should detect these attacks in a short time.
To evaluate this, we design a masquerading attack where the adversary not only knows the password but also observes and mimics the user's behavioral patterns. If the adversary succeeds in mimicking
the user's behavioral pattern, then SmarterYou will misidentify the adversary as the legitimate user and he/she can thus use the victim user's smartphone. 

\begin{figure}[!t] \centering
\includegraphics[width=2in,height=1.1in]{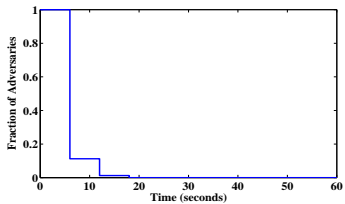}
\DeclareGraphicsExtensions.
\caption{Fraction of adversaries that have access to the legitimate user's smartphone at time $t$.}
\label{fig:fraction}
\end{figure}

In order to show the ability of SmarterYou to defend against these mimicry attacks, we counted the percentage of people (attackers) who were still using the smartphone without being de-authenticated by the system as the attack time progresses. Figure \ref{fig:fraction} shows the fraction of adversaries that are recognized as legitimate users by SmarterYou at time $t$, from which we can see how quickly SmarterYou can recognize an adversary and terminate his access to the smartphone. At $t=0$, all the adversaries have access to the smartphone, but within $6$s, only $10\%$ of adversaries have access. That is, SmarterYou identified on average $90\%$ of adversaries as unauthorized users within $6$s. By $t=18$s, SmarterYou identified all the adversaries. Therefore, SmarterYou performed well in recognizing the adversary who is launching the masquerading attack. 

These experimental results also match with analysis from a theoretical point of view. We assume the FAR in each time window is $p$, then the chance that the attacker can escape from detection in $n$ time windows is $p^n$. Based on our experimental results in Section \ref{sec:authentication}, our system can achieve $2.8\%$ FAR in a time window of 6 seconds. Thus, within only three windows, the probability for the attacker escaping detection is $(2.8\%)^3 = 0.002\%$, which is very small.  Therefore, our SmarterYou system shows good performance in defending against masquerading attacks.

\subsection{Smartphone Overhead}\label{sec:overhead}
We now evaluate the system overhead of SmarterYou on smartphones. 
Specifically, we analyze the computational complexity of our system, CPU and memory overhead, and the battery consumption it incurs on the smartphone.
\subsubsection{Computational Complexity}
The computational complexity of KRR in Section~\ref{sec:krr} is directly related to the data size according to Eq.~\ref{optimal}. Here, we further show that the computational complexity can be largely reduced to be directly related to the feature size. \new{(For readability, we put the detailed proof in the Appendix)}.

\noindent According to Eq.~\ref{optimal}, the classifier is $\bm{w}^{*} =\bm{\Phi}[\bm{K}+\rho \bm{I_N}]^{-1} \bm{y}$.

Define $\bm{S}=\bm{\Phi\Phi}^T$ ($\bm{\Phi} = [\bm{\vec{\phi}(\bm{x_1})},\bm{\vec{\phi}(\bm{x_2})},\cdots, \bm{\vec{\phi}(\bm{x_N})}]$). By utilizing the matrix transformation method in \cite{horn2012matrix}, the optimal solution $\bm{w}^{*}$ in Eq.~\ref{optimal} is equivalent to
\begin{equation}\label{optimal2}
\bm{w}^{*} =[\bm{S}+\rho \bm{I_J}]^{-1}\Phi \bm{y}
\end{equation}
The dominant computational complexity for $\bm{w}^{*}$ comes from taking the inversion of a matrix. Therefore, based on Eq.~\ref{optimal} and Eq.~\ref{optimal2}, the computational complexity is approximately $\min (O(N^{2.373}), O(J^{2.373}))$. If we utilize the identity kernel, the computational complexity can be reduced from $O(N^{2.373})$ to $O(M^{2.373})$ and is 
independent of the data size. Specifically, we construct $28$-dimensional feature vectors ($4$ time-domain features and $3$ frequency-domain features for each of two sensors, for each device).

Thus, our time complexity is reduced from $O((800\times 9/10)^{2.373})=O(720^{2.373})$ to only $O(28^{2.373})$. In our experiments, the average training time is $0.065$ seconds and the average testing time is $18$ milliseconds, which shows the effectiveness of our system applied in real-world scenarios. 

\begin{figure}[!t]
\centering
\includegraphics[width=3.2in,height=1.1in]{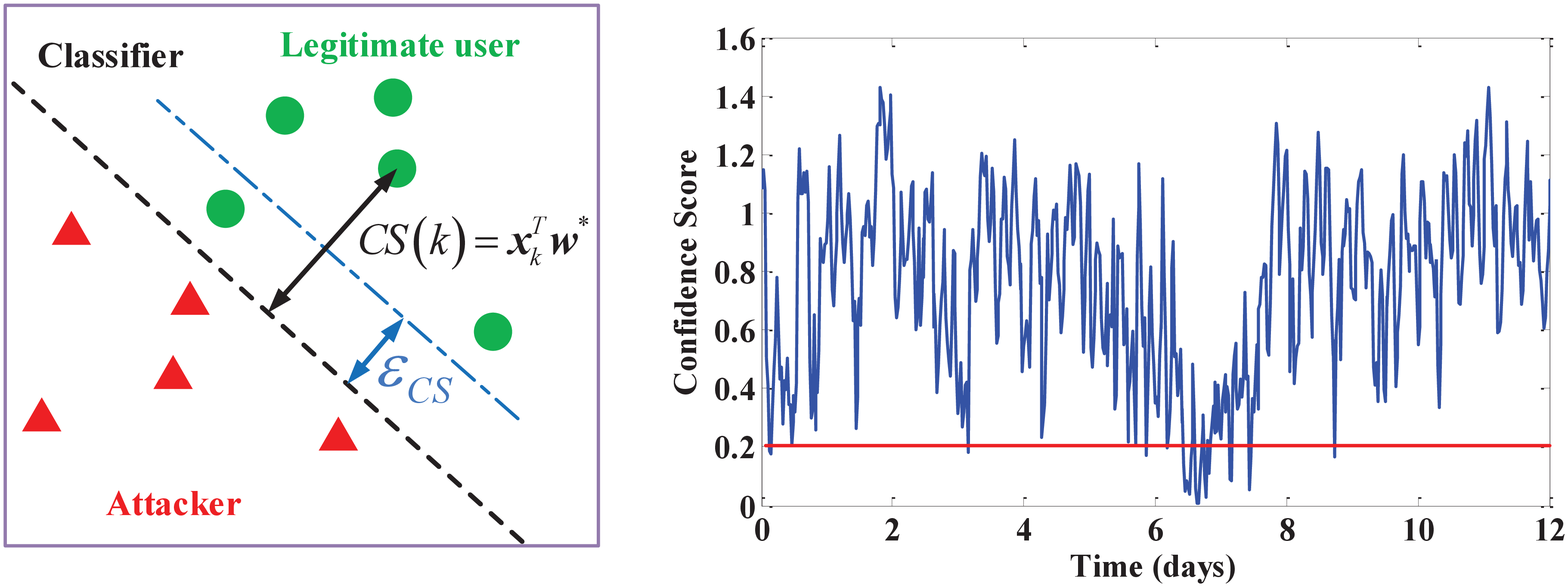}
\DeclareGraphicsExtensions.
\caption{The confidence score of a user with time. After around one week, the confidence score decreases below the threshold $\epsilon_{CS}=0.2$ for a period of time. After automatic retraining, it increases back to normal values.}
\label{fig:re-train}
\end{figure}

\subsubsection{CPU and Memory Overhead}
The testing module of SmarterYou in a smartphone runs as threads inside the smartphone system process. We develop an application to monitor the average CPU and memory utilization of the phone and watch while running the SmarterYou app which continuously requests sensor data at a rate of $50$ Hz on a Nexus 5 smartphone and a Moto 360 smartwatch. The CPU utilization is $5\%$ on average and never exceeds $6\%$. The CPU utilization (and hence energy consumption) will scale with the sampling rate. The memory utilization is $3$ MB on average. Thus, we believe that the overhead of SmarterYou is small enough to have negligible effect on overall smartphone performance.

\subsubsection{Battery Consumption}
To measure the battery consumption, we consider the following four testing scenarios: (1) Phone is locked (i.e., not being used) and SmarterYou is off. (2) Phone is locked and SmarterYou keeps running.  (3) Phone is under use and SmarterYou is off. (4) Phone is under use and SmarterYou is running. For scenarios (1) and (2), the test time is 12 hours each. We charge the smartphone battery to $100\%$ and check the battery level after 12 hours. The average difference of the battery charged level from $100\%$ is reported in Table \ref{table:power}.

For scenarios (3) and (4), \emph{the phone under use} means that the user keeps using the phone periodically. During the using time, the user keeps typing notes. The period of using and non-using is five minutes each, and the test time in total is 60 minutes.

Table \ref{table:power} shows the result of our battery consumption tests, 
in terms of extra battery drain for SmarterYou. We find that in scenarios (1) and (2), the SmarterYou-on mode consumes $2.1\%$ more battery power than the SmarterYou-off mode. We believe the extra cost in battery consumption caused by SmaterYou will not affect user experience in daily use. For scenarios (3) and (4), SmarterYou consumes $2.4\%$ more battery power in one hour, which is also an acceptable cost for daily usage.

\begin{table}[!t]\scriptsize
\centering
\caption{The power consumption under four different scenarios.}
\begin{tabular}{|l|c|} \hline
Scenario & \tabincell{c}{Power Consumption} \\ \hline
(1) Phone locked, SmarterYou off & $2.8\%$ \\ \hline
(2) Phone locked, SmarterYou on & $4.9\%$ \\ \hline
(3) Phone unlocked, SmarterYou off & $5.2\%$ \\ \hline
(4) Phone unlocked, SmarterYou on & $7.6\%$ \\ \hline
\end{tabular}
\label{table:power}
\end{table} 

\subsection{Retraining Authentication Models}
\label{sec:retrain_model}
The behavioral drift of the legitimate user must be considered.  The user may change his/her behavioral pattern over weeks or months, which may cause more false alarms in implicit authentication. SmarterYou, therefore, will retrain the authentication models automatically and continuously based on the previous authentication performance. Here, we define the confidence score (CS) \ruby{as $CS(k)={\bm{x}}_k^T\bm{w}^{*}$} for the $k$-th authentication feature vector ${\bf{x}}_k^T$ as the distance between ${\bf{x}}_k^T$ and the corresponding authentication classifier $\bm{w}^{*}$.

As the authentication classifier $\bm{w}^{*}$ represents the classification boundary to distinguish the legitimate user and the adversaries, a lower confidence score (smaller distance between ${\bf{x}}_k^T$ and $\bm{w}^*$) represents a less confident authentication result (shown conceptually in the left figure of Figure~\ref{fig:re-train}). This suggests a change of user's behavioral pattern where retraining should be taken. For an authenticated user, we suggest that if the confidence score is lower than a certain threshold $\epsilon_{CS}$ for a period of time $T$, the system automatically retrains the authentication models.

In Figure~\ref{fig:re-train} (right figure), we show the confidence score of the time-series authentication feature vectors for a user. We can see that the confidence score decreases slowly in the first week. At the end of the first week, the confidence score experiences a period of low values (lower than our threshold $\epsilon_{CS}=0.2$ for a period), indicating that the user's behavior changes to some extent during this week. Therefore, it would be helpful if the system can automatically 
retrain the authentication models. Note that there are some earlier points lower than the threshold ($0.2$), but they do not occur for a long enough period to trigger the retraining. \hank{Also, it is hard for the attacker to trigger the retraining because the probability that the attacker continuously passes the authentication for a long period of time is low as described in Section \ref{sec:security}.}

As our system recognizes user's behavior drift by checking the confidence score, it would then go back to the training module again and upload the legitimate user's authentication feature vectors to the training module until the new behavior (authentication model) is learned. Advanced approaches in machine unlearning~\cite{cao2015towards} can be explored to update the authentication models asymptotically faster than retraining from scratch. After retraining the user's authentication models, we can see that the confidence score increases to normal values from Day $8$.

As discussed earlier, an attacker who has taken over a legitimate user's smartphone must not be allowed to retrain the authentication model. Fortunately, the attacker can not trigger the retraining since the confidence score should be positive and last for a period of time. However, the attacker is likely to have negative confidence scores, which cannot last for sufficient time to trigger retraining, since he will be detected in less than $18$ seconds by SmarterYou, according to Figure~\ref{fig:fraction}.

\section{Conclusions}\label{sec:conclusion}
We have proposed a new re-authentication system, SmarterYou, to improve the security of a smartphone, and of secret and sensitive data and code in the smartphone or in the cloud accessible through a smartphone. 
SmarterYou is an authentication system using multiple sensors built into a user's smartphone, supplemented by auxiliary information from a wearable device, e.g., smartwatch, with the same owner as the smartphone. Our system keeps monitoring the users' sensor data and continuously authenticates without any human cooperation. We first collect context features from the sensors' data in the smartphone (and the smartwatch if present) to detect the context of the current user. Based on the detected context and the authentication features in both the time and frequency domains, our system implements finer-grained authentication efficiently and stealthily. 

We systematically evaluate design alternatives for each design parameter of such a sensor-based implicit authentication system.
Based on our design choices, our evaluations demonstrate the advantage of combining the smartphone and the smartwatch and the enhancement in authentication accuracy with context detection and time-frequency information. SmarterYou can achieve authentication accuracy up to $98.1\%$ (FRR $0.9\%$ and FAR $2.8\%$) with negligible system overhead and less than $2.4\%$ additional battery consumption. We believe this is the highest accuracy and lowest FAR reported by any sensor-based authentication method to date. We hope that the SmarterYou system and design techniques can help advance the field in implicit user authentication and re-authentication, for deployment in real-world scenarios.

\bibliographystyle{IEEEtran}
\bibliography{v3}  

\section{Appendix}\footnotesize
\subsection{Proof of Equivalence between Eq.~\ref{optimal} and Eq. \ref{optimal2}}

\noindent Eq.~\ref{optimal} is $\bm{w}^{*} =\bm{\Phi}[\bm{K}+\rho \bm{I_N}]^{-1} \bm{y}$,  and Eq. \ref{optimal2} is $\bm{w}^{*} =[\bm{S}+\rho \bm{I_J}]^{-1}\Phi \bm{y}$.
In order to prove that they are equivalent, we first prove $\bm{PB^T(BPB^T}+\rho \bm{R})^{-1} = (\rho \bm{P^{-1}+B^TR^{-1}B)^{-1}B^TR^{-1}}$ as follows:
\begin{equation}\label{final}
\begin{aligned}
& \rho \bm{B^T+ B^TR^{-1}BPB^T} = \bm{B^TR^{-1}BPB^T+}\rho \bm{B^T} \\
& \Leftrightarrow (\rho \bm{P^{-1}+ B^TR^{-1}B)PB^T} = \bm{B^TR^{-1}(BPB^T+}\rho \bm{R)}\\
& \Leftrightarrow \bm{PB^T(BPB^T+}\rho \bm{R)^{-1}} = (\rho \bm{P^{-1}+B^TR^{-1}B)^{-1}B^TR^{-1}}\\
\end{aligned}
\end{equation}

Then we let $\bm{P =I_J}$, $\bm{B= \Phi}^T$ and $\bm{R= I_N}$ in Eq. \ref{final}, we observe the left hand side of Eq. \ref{final} is Eq.~\ref{optimal} and the right hand side of Eq. \ref{final} is Eq. \ref{optimal2}. Thus, we prove the equivalence between Eq.~\ref{optimal} and Eq. \ref{optimal2}.

\end{document}